\documentclass[oneside,a4paper,11pt,shownumbers]{article}
\usepackage{latexsym}
\usepackage{euscript}
\usepackage{epsfig,amsmath,amssymb}

\def\be{\begin{equation}}
\def\ee{\end{equation}}
\def\bea{\begin{eqnarray}}
\def\eea{\end{eqnarray}}

\long\def\symbolfootnote[#1]#2{\begingroup%
\def\thefootnote{\fnsymbol{footnote}}\footnote[#1]{#2}\endgroup} 

 \large\normalsize

\begin{document}

\begin{center}

{\Large \bf Symmetry breaking in (gravitating) scalar field models describing interacting boson stars and $Q$-balls}

\vspace*{7mm} {Yves Brihaye $^{a}$
\symbolfootnote[1]{E-mail: yves.brihaye@umh.ac.be},
Thierry Caebergs $^{a}$
\symbolfootnote[2]{E-mail:thierry.caebergs@umh.ac.be},
Betti Hartmann $^{b}$
\symbolfootnote[3]{E-mail:b.hartmann@jacobs-university.de}
and Momchil Minkov $^{b}$
\symbolfootnote[4]{E-mail:m.minkov@jacobs-university.de}
}
\vspace*{.25cm}

${}^{a)}${\it Facult\'e des Sciences, Universit\'e de Mons-Hainaut, 7000 Mons, Belgium}\\
${}^{b)}${\it School of Engineering and Science, Jacobs University Bremen, 28759 Bremen, Germany}\\

\vspace*{.3cm}
\end{center}

\begin{abstract}
We investigate the properties of interacting $Q$-balls and boson stars that sit on top of each other in great detail. The model that describes these solutions is essentially
a (gravitating) two-scalar field model where both scalar fields are complex.
We construct interacting $Q$-balls or boson stars
with arbitrarily small charges but finite mass.
We observe that in the interacting case --
where the interaction can be either due to the potential or due to gravity -- two types of solutions
exist for equal frequencies: one for which
the two scalar fields are equal, but also one for which the two scalar fields differ.
This constitutes a symmetry breaking in the model. While for $Q$-balls asymmetric solutions
have always corresponding symmetric solutions and are thus likely unstable to decay to symmetric
solutions with lower energy, there exists a parameter regime for interacting boson stars, where only asymmetric
solutions exist. 
We present the domain of existence for two interacting non-rotating solutions as well
as for solutions describing the interaction between rotating and non-rotating $Q$-balls and boson
stars, respectively.

\end{abstract}

\section{Introduction}
Solitons play an important role in many areas of physics. As classical solutions of non-linear field theories, they
are localized structures with finite energy, which are globally regular.
In general, one can distinguish between topological and non-topological solitons.
While topological solitons \cite{ms} possess a conserved quantity, the topological charge, that stems (in most
cases) from the spontaneous symmetry breaking of the theory, non-topological solitons \cite{fls,lp} have a conserved Noether
charge that results from a symmetry of the Lagrangian. The standard example of  non-topological solitons
are $Q$-balls \cite{coleman}, which are solutions of theories with self-interacting complex scalar fields. These objects are stationary with an explicitly
time-dependent phase. The conserved Noether charge $Q$
is then related to the global phase invariance of the theory and is directly proportional
to the frequency. $Q$ can e.g. be interpreted as particle number \cite{fls}. 

While in standard scalar
field theories, it was shown
that a non-renormalisable $\Phi^6$-potential is necessary \cite{vw}, supersymmetric extensions of the
Standard Model (SM)  also possess $Q$-ball solutions \cite{kusenko}. In the latter case, several scalar fields
interact via complicated potentials. It was shown that cubic interaction terms that result from
Yukawa couplings in the superpotential and supersymmetry breaking terms lead to the existence of $Q$-balls
with non-vanishing baryon or lepton number or electric charge. These supersymmetric
$Q$-balls have been considered as possible candidates for baryonic dark matter 
\cite{dm} and their astrophysical implications have been discussed \cite{implications}.
In \cite{cr}, these objects have been constructed numerically using 
the exact form of the supersymmetric potential.

$Q$-ball solutions in $3+1$ dimensions have been studied in detail in 
\cite{vw,kk1,kk2}. It was realized
that next to non-spinning $Q$-balls, which are spherically symmetric, spinning solutions
exist. These are axially symmetric with energy density of toroidal shape
and angular momentum $J=kQ$, where $Q$ is the Noether charge of the solution
and $k\in \mathbb{Z}$ corresponds to the winding around the $z$-axis. 
Approximated  solutions of the non-linear partial differential equations
were constructed in \cite{vw} by means of a truncated series in the spherical harmonics to describe
the angular part of the solutions. 
The full  partial differential equation was solved numerically in \cite{kk1,kk2,bh}. 
It was also realized in \cite{vw} that in each $k$-sector, parity-even ($P=+1$)
and parity-odd ($P=-1$) solutions exist. Parity-even and parity-odd 
refers to the fact that
the solution is symmetric and anti-symmetric, respectively with respect
to a reflection through the $x$-$y$-plane, i.e. under $\theta\rightarrow \pi-\theta$.

These two types of solutions are
closely related to the fact that the angular part of the solutions
constructed in \cite{vw,kk1,kk2}
is connected to the spherical harmonic $Y_0^0(\theta,\varphi)$ for the spherically symmetric $Q$-ball,
to the spherical harmonic $Y_1^{1}(\theta,\varphi)$ for the spinning parity even ($P=+1$) solution
and to the spherical harmonic $Y_2^{1}(\theta,\varphi)$  for the parity  odd ($P=-1$) solution, respectively. 
Radially excited solutions of the spherically symmetric, non-spinning solution were also obtained.
These solutions are still spherically symmetric but the scalar field develops one or several nodes for $r\in ]0,\infty[$. 
In relation to the apparent connection of the angular part of the known solutions to the spherical harmonics,
``$\theta$-angular excitations'' of the $Q$-balls 
corresponding to the spherical harmonics  $Y_l^k(\theta,\varphi)$, $-l \leq k \leq l$ have been constructed
explicitely for some values of $k$ and $l$ in \cite{bh}.
These excited solutions could play a role in the formation of $Q$-balls in the early universe
since it is believed that $Q$-balls forming due to condensate fragmentation at the end of inflation first appear in an excited state and only then settle down to the ground state \cite{nonspherical}. The fact that these newly formed $Q$-balls are excited, i.e.
in general not spherically symmetric could, on the other hand,  be a source of gravitational waves \cite{km2}.

Interactions of well-separated $Q$-balls in $(1+1)$-dimensions have been studied in \cite{bfs} and it was
shown that $Q$-balls with equal frequencies can attract when being in-phase or repell when being exactly
out-of-phase.

The interaction of two $Q$-balls in $(3+1)$-dimensions that sit on top of each other has been studied in \cite{bh}.
It was found that the lower bound on the frequencies $\omega_i$, $i=1,2$
is increasing for increasing interaction coupling. 
Explicit examples of a rotating $Q$-ball interacting with a non-rotating
$Q$-ball have been presented.

Complex scalar field models coupled to gravity possess so-called ``boson star'' solutions \cite{kaup,misch,flp,jetzler,new1,new2}.
In \cite{kk1,kk2,bh2} boson stars have
been considered that have flat space-time limits in the form of
$Q$-balls. These boson stars are hence self-gravitating $Q$-balls. The interaction of boson stars has been studied in \cite{bh2} and it was found that ergoregions
can appear when a non-rotating boson star interacts with a rotating and parity even boson star
signaling an instability of the solution.
Recently, charged Q-balls and boson stars in scalar electrodynamics have also been considered \cite{kkll}.

In this paper, we study interacting boson stars and $Q$-balls that sit on top of each other.
While in \cite{bh,bh2} we were mainly interested in the different types of solutions
existing in the model, we present here an analysis of the dependence of charges and masses
of the solutions on the parameters of the model. During this analysis, we have found that the pattern
of solutions is much richer than in the non-interacting case, 
especially we observe new (and unexpected) branches
specific to the system of two scalar fields. Fixing the different
coupling constants appearing in the Lagrangian in such a way that it is
symmetric under the exchange of the two scalar fields,
we impose an extra $Z_2$ symmetry.
Then, it turn out that the new branches 
correspond to solutions which break this symmetry. In other
words, the $Z_2$ symmetry is spontaneously broken by the new solutions.

The paper is organized as follows: in Section 2, we give the model, Ansatz and boundary conditions. In Section 3 and 4, we discuss our results for non-rotating and rotating solutions, respectively. 
Section 5 contains our conclusions.

\section{The model}
In the following, we study a scalar field model coupled minimally to gravity in $3+1$ dimensions describing two interacting boson stars. 
The action $S$ reads:
\begin{equation}
 S=\int \sqrt{-g} d^4 x \left( \frac{R}{16\pi G} + {\cal L}_{m}\right)
\end{equation}
where $R$ is the Ricci scalar, $G$ denotes Newton's constant and ${\cal L}_{m}$ denotes
the matter Lagrangian: 
\begin{equation}
\label{lag}
 {\cal L}_{m}=-\frac{1}{2}\partial_{\mu} \Phi_1 \partial^{\mu} \Phi_1^*-
\frac{1}{2}\partial_{\mu} \Phi_2 \partial^{\mu} \Phi_2^* - V(\Phi_1,\Phi_2)
\end{equation}
where both $\Phi_1$ and $\Phi_2$ are complex scalar fields and we choose as signature of the metric
$(-+++)$. The potential reads:
\begin{equation}
V(\Phi_1,\Phi_2)=\sum_{i=1}^2\left(
\kappa_i \vert\Phi_i\vert^6 - \beta_i \vert\Phi_i\vert^4 +
\lambda_i \vert\Phi_i\vert^2 \right)
+\gamma \vert\Phi_1\vert^2 \vert\Phi_2\vert^2
\end{equation} 
where $\kappa_i$, $\beta_i$, $\lambda_i$, $i=1,2$ are the standard potential parameters for each boson star, while $\gamma$ denotes the interaction parameter. The masses of the two bosonic scalar fields are then given by $(m_B^i)^2 = \lambda_i$, $i=1,2$.

Along with
\cite{kk1,kk2,bh}, we choose in the following 
\begin{equation}
\label{parameters}
 \kappa_i=1 \ \ , \ \ \beta_i=2 \ \ , \ \ \lambda_i=1.1 \ \ , \ \ i=1,2 \ .
\end{equation}
This particular choice of parameters leads to an extra $Z_2$ symmetry of the Lagrangian: $\Phi_1\leftrightarrow \Phi_2$.

In \cite{vw} it was argued that a $\Phi^6$-potential is necessary in order to have classical
$Q$-ball solutions. This is still necessary for the model we have defined here,
since we want $\Phi_1=0$
and $\Phi_2=0$ to be a local minimum of the potential. A pure $\Phi^4$-potential which is bounded from below wouldn't fulfill these criteria. 
 
The matter Lagrangian ${\cal L}_{m}$ (\ref{lag}) is invariant under the two independent global U(1) transformations
\begin{equation}
 \Phi_1 \rightarrow \Phi_1 e^{i\chi_1} \ \ \ , \ \ \ 
\Phi_2 \rightarrow \Phi_2 e^{i\chi_2}   \ .
\end{equation}
As such the total conserved Noether
current $j^{\mu}_{(tot)}$, $\mu=0,1,2,3$, associated to these symmetries is just the sum of
the two individually conserved currents $j^{\mu}_{1}$ and $j^{\mu}_2$ with
\begin{equation}
j^{\mu}_{(tot)}= j^{\mu}_1 +j^{\mu}_2
 = -i \left(\Phi_1^* \partial^{\mu} \Phi_1 - \Phi_1 \partial^{\mu} \Phi_1^*\right)
-i  \left(\Phi_2^* \partial^{\mu} \Phi_2 - \Phi_2 \partial^{\mu} \Phi_2^*\right)\ \ .
\end{equation}
with $j^{\mu}_{1\ ;\mu}=0$, $j^{\mu}_{2 \ ;\mu}=0$ and  $j^{\mu}_{(tot) \ ; \mu}=0$.

The total Noether charge $Q_{(tot)}$ of the system is then the sum of the two individual Noether charges $Q_1$ and $Q_2$:
\begin{equation}
 Q_{(tot)}=Q_1+Q_2= -\int j_1^0 d^3 x  -\int j_2^0 d^3 x
\end{equation}

Finally, the energy-momentum tensor reads:
\begin{equation}
\label{em}
T_{\mu\nu}=\sum_{i=1}^2 \left(\partial_{\mu} \Phi_i \partial_{\nu} \Phi_i^*
+\partial_{\nu} \Phi_i \partial_{\mu} \Phi_i^*\right) -g_{\mu\nu} {\cal L}
\end{equation}

\subsection{Ansatz and Equations}
For the metric the Ansatz in Lewis-Papapetrou form reads \cite{kk1}:
\begin{equation}
 ds^2 = - fdt^2 + \frac{l}{f}\left(g (dr^2 + r^2 d\theta^2) + r^2 \sin^2 \theta (d\varphi + \frac{m}{r} dt)^2 \right)
\end{equation}
where the metric functions $f$, $l$, $g$ and $m$ are functions of $r$ and $\theta$ only.
For the scalar fields, the Ansatz reads:
\begin{equation}
\label{ansatz1}
\Phi_i(t,r,\theta,\varphi)=e^{i\omega_i t+ik_i\varphi} \phi_i(r,\theta) \  \ ,
\ i=1,2
\end{equation}
where the $\omega_i$ and the $k_i$ are constants. Since we require $\Phi_i(\varphi)=\Phi_i(\varphi+2\pi)$, $i=1,2$, we have $k_i\in \mathbb{Z}$.
The mass $M$ and total angular momentum $J$ of the solution can be read off from the asymptotic behaviour
of the metric functions \cite{kk1}:
\begin{equation}
 M=\frac{1}{2G} \lim_{r\rightarrow \infty} r^2 \partial_r f  \ \ , \ \
J=\frac{1}{2G} \lim_{r\rightarrow \infty} r^2 m   \ .
\end{equation}
The total angular momentum $J=J_1+J_2$ and the Noether charges $Q_1$ and $Q_2$ of the two boson stars are related by $J=k_1 Q_1 + k_2 Q_2$. Boson stars with $k_i=0$ have thus
vanishing angular momentum. Equally, interacting boson stars with $k_1=-k_2$ and $Q_1=Q_2$ have vanishing
angular momentum. 

The coupled system of partial differential equations is then given by the Einstein
equations
\begin{equation}
\label{einstein}
 G_{\mu\nu}=8\pi G T_{\mu\nu}
\end{equation}
with $T_{\mu\nu}$ given by (\ref{em}) and the Klein-Gordon equations 
\begin{equation}
\label{KG}
 \left(\square + \frac{\partial V}{\partial \vert\Phi_i\vert^2} \right)\Phi_i=0 \ \ , \ \ i=1,2 \ .
\end{equation}
The explicit expressions for the equations can be found in the Appendix.

So far, the scale of the scalar fields is not yet fixed and 
the functions $\phi_1$, $\phi_2$ are dimensionful. In order to study
the equations, it is convenient to rescale these function
according to
\begin{equation}
 \phi_i \to \phi_i\eta  \ \ , \ \ i=1,2   \ .
\end{equation}
where $\eta$ has the dimension of an energy and its scale
depends on the chosen phenomenological model \cite{kusenko,dm,implications,cr}.
The potential parameters are rescaled as $\gamma \to \gamma/\eta^2$, $\beta_i\to \beta_i/\eta^2$
and $\kappa_i\to \kappa_i/\eta^4$.

We can then introduce the dimensionless quantity
\begin{equation}
 \alpha=8\pi G\eta^2=\frac{8\pi\eta^2}{M_{pl}^2}
\end{equation}
which measures the ratio between the energy scale of the scalar field and the Planck mass.

As stated in \cite{cr}, in supersymmetric extensions of the standard model, 
the value of $\eta$ is between a few TeV and a few hundred TeV.
The value of $\alpha$ is hence very small. However, in order to be able to 
compare our results with the case of a single $Q$-ball or boson star, we follow
the discussion in \cite{kk1,kk2} and also study larger values of $\alpha$
throughout this paper. Furthermore, larger values of $\alpha$ also accommodate
possible other energy scales.

\subsection{Boundary conditions}

We require the solutions to be regular at the origin. The appropriate boundary conditions
read:
\begin{equation}
\partial_r f|_{r=0}=0 \ , \ \ \
\partial_r l|_{r=0}=0 \ , \ \ \
g|_{r=0}=1 \ , \ \ \
m|_{r=0}=0 \ , \ \ \
\phi_i| _{r =0}=0 \  \ , \ \ i=1,2.
\label{bc3} \end{equation}
for solutions with $k_i\neq 0$, while for $k_i=0$ solutions, we have $\partial_r \phi_i|_{r=0}=0$, $i=1,2$.
The boundary conditions at infinity result from the requirement of asymptotic
flatness and finite energy solutions: 
\begin{equation}
f|_{r \rightarrow \infty} =1 \ , \ \ \
l|_{r \rightarrow \infty} =1 \ , \ \ \
g|_{r \rightarrow \infty} =1 \ , \ \ \
m|_{r \rightarrow \infty} =0 \ , \ \ \
\phi_i| _{r \rightarrow \infty}=0 \ \ , \ \ i=1,2.
\label{bc4} \end{equation}

For $\theta=0$ the regularity of the solutions on the $z$-axis requires:
\begin{equation}
\partial_{\theta} f|_{\theta=0}=0 \ , \ \ \
\partial_{\theta} l|_{\theta=0}=0 \ , \ \ \
g|_{\theta=0}=1 \ , \ \ \
\partial_{\theta} m |_{\theta=0}=0 \ , \ \ \
\phi_i |_{\theta=0}=0 \ \ , \ \ i=1,2 \ , 
\label{bc5} \end{equation}
for $k_i\neq 0$ solutions, while for $k_i=0$ solutions, we have $\partial_{\theta} \phi_i|_{r=0}=0$, $i=1,2$.

The conditions at $\theta=\pi/2$
are either given by
\begin{equation}
\partial_{\theta} f|_{\theta=\pi/2}=0 \ , \ \ \
\partial_{\theta} l|_{\theta=\pi/2}=0 \ , \ \ \
\partial_{\theta} g|_{\theta=\pi/2}=0 \ , \ \ \
\partial_{\theta} \omega |_{\theta=\pi/2}=0 \ , \ \ \
\partial_{\theta} \phi_i |_{\theta=\pi/2}=0 \ \ , \ \ i=1,2
\label{bc6} \end{equation}
for even parity solutions,
while for odd parity solutions the conditions for the scalar field functions read:
$\phi_i |_{\theta=\pi/2}=0$, $i=1,2$.

\section{Non-rotating solutions}
The solutions have vanishing angular momentum for $k_1=k_2=0$. 
In this case, the system of differential equations
reduces to a system of coupled ordinary differential equations and
$g\equiv 1$, $m\equiv 0$ and the remaining functions are functions of the radial variable
$r$ only.

We have solved the corresponding ordinary differential equations (ODEs) using the ODE solver
COLSYS \cite{colsys}.

\subsection{$\alpha=0$~: Interacting $Q$-balls}
In the flat space-time limit, i.e. for $\alpha=0$ the metric functions
$f=l\equiv 1$ and the system describes two interacting, non-rotating $Q$-balls. 
These $Q$-balls are interacting only if $\gamma\neq 0$ via the potential interaction.
For single $Q$-balls, it has been observed \cite{vw,kk1,kk2} that the solutions exist
only on a finite interval of the frequency $\omega$. It was realized in \cite{bh} that
this is also true for interacting $Q$-balls. In the limit of two non-rotating $Q$-balls the upper bounds on $\omega_1$ and $\omega_2$ are \cite{bh}:
\begin{equation}
\label{omegamax}
 \omega^2_1 \leq \omega^2_{1,max} = \lambda_1 \ \ , \ \
\omega^2_2 \leq \omega^2_{2,max} = \lambda_2 
\end{equation}
To find the lower bound, we introduce a polar decomposition of $\phi_1$ and $\phi_2$:
\begin{equation}
 \phi_1 = \rho \cos\chi \ \ , \ \ \phi_2=\rho \sin\chi
\end{equation}
such that the lower bound reads \cite{bh}:
\begin{eqnarray}
\label{lower_bound}
 \omega_1^2 \cos^2\chi &+& \omega_2^2\sin^2\chi  \geq  \left(\omega_1^2 \cos^2\chi + \omega_2^2\sin^2\chi\right)_{min} \nonumber \\
&=& \lambda_1 \cos^2\chi + \lambda_2 \sin^2\chi - \frac{1}{4} \frac{\left(\beta_1 \cos^4\chi + \beta_2 \sin^4 \chi - \gamma \cos^2\chi \sin^2\chi\right)^2}{\kappa_1 \cos^6\chi + \kappa_2 \sin^6\chi}
\end{eqnarray}

We have studied the dependence of the charges and the mass of the solution in dependence
on the potential parameter $\gamma$. We have also investigated how the mass
of the solution evolves in comparison to the mass of $Q_1 + Q_2$ individual bosons with masses
$m_{B,1}$ and $m_{B,2}$, respectively.

The pattern of solutions is involved when studying the dependence of the solutions on $\omega_1$, $\omega_2$ and $\gamma$. Thus, we first discuss the case $\omega_1=\omega_2$, where we expect in analogy to the non-interacting
case with $\gamma=0$ that we should always find $\phi_1=\phi_2$, i.e. $Q_1=Q_2$. However, already in this limit,
the pattern of solutions turns out to be richer than expected. We come back to this observation
at the end of this subsection. First, let us discuss the effect of $\gamma$ on solutions with $Q_1=Q_2$.

\subsubsection{$\omega_1=\omega_2$}
\begin{figure}[!htb]
\centering
\leavevmode\epsfxsize=15.0cm
\epsfbox{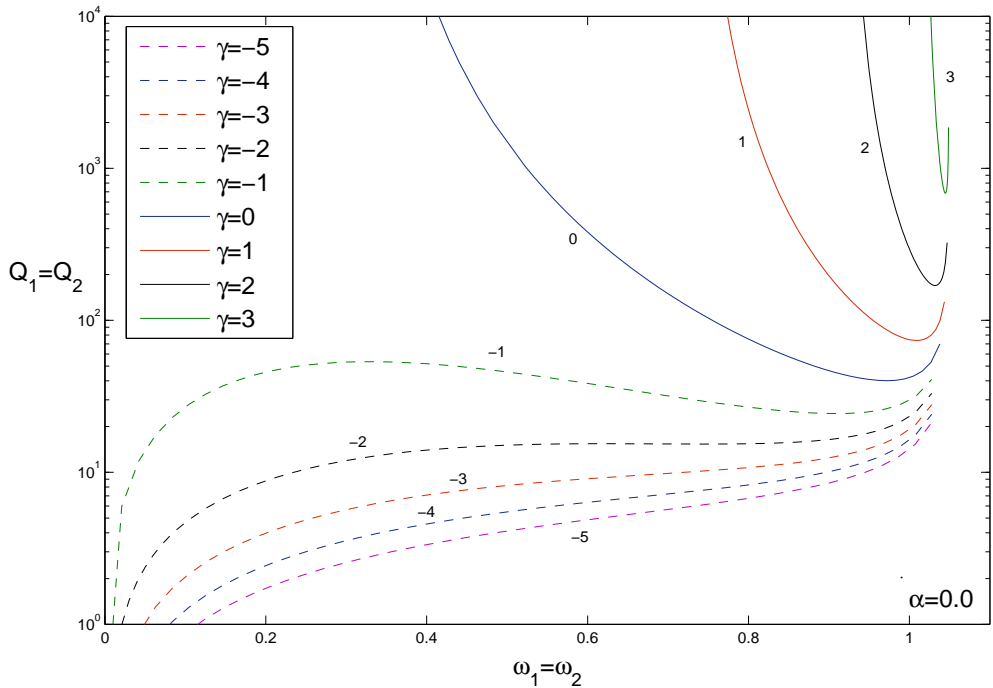}\\
\caption{\label{alp0q} The dependence of the charges $Q_1=Q_2$ of two interacting $Q$-balls on $\omega_1=\omega_2$ is shown for  different values of $\gamma$. The small numbers indicate
the value of $\gamma$. }
\end{figure}

\begin{figure}[!htb]
\centering
\leavevmode\epsfxsize=15.0cm
\epsfbox{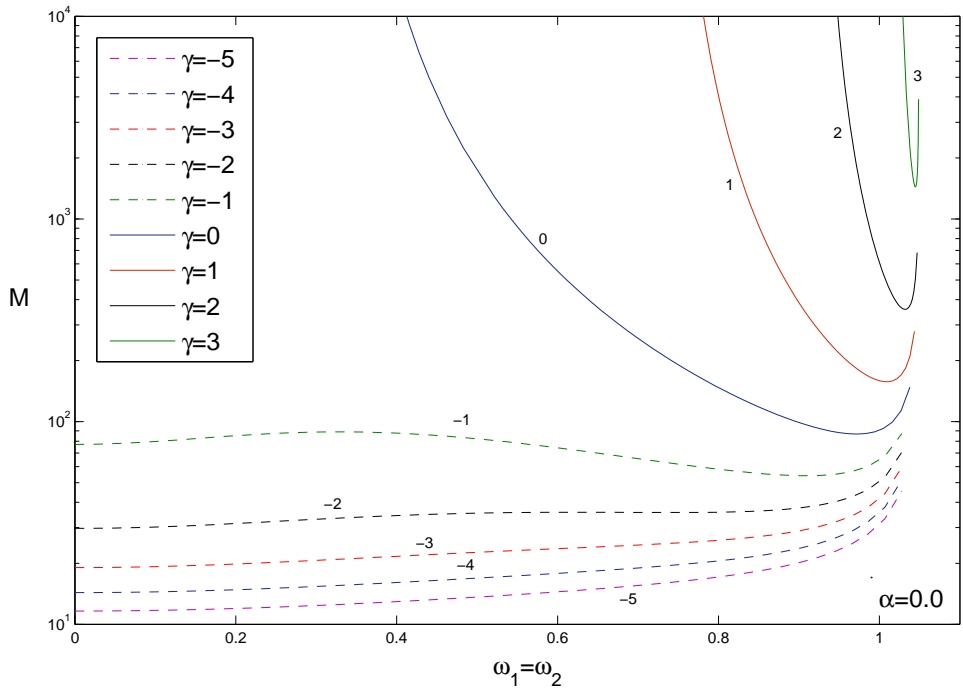}\\
\caption{\label{alp0m} The dependence of the mass $M$ of two interacting $Q$-balls on $\omega_1=\omega_2$ is shown for  different values of $\gamma$. The small numbers indicate
the value of $\gamma$. }
\end{figure}

We first discuss the case $\phi_1=\phi_2$, i.e. $\chi=\pi/4$. Then the 
bounds on the frequencies are given by (\ref{omegamax}) and (see (\ref{lower_bound})):
\begin{equation}
\label{omegamin}
\left(\omega_1^2 + \omega_2^2\right)_{min}=  \frac{1}{5} + \gamma - \frac{1}{8}\gamma^2 \leq \omega_1^2 + \omega_2^2 
\end{equation}
For $\omega_1=\omega_2$ tending to the upper bound (\ref{omegamax}) the charges $Q_1=Q_2$ and the mass $M$ diverge. This is clearly seen in Figs.\ref{alp0q} and \ref{alp0m}, where we plot the
dependence of the charges $Q_1=Q_2$ and the mass $M$ of the solution on $\omega_1=\omega_2$
for different values of $\gamma$. Apparently the charges and the mass diverge for
$\omega_1\rightarrow \sqrt{\lambda_1} \approx 1.0488$.

For $\omega_1=\omega_2$ tending to the lower bound (\ref{omegamin}), we observe that the limiting behaviour depends crucially on the value of $\gamma$. To understand this, we observe that the lower bound (\ref{omegamin}) becomes negative for $\gamma \leq 4 - \sqrt{88/5} \approx -0.195$ and $ \gamma \geq 4 + \sqrt{88/5} \approx 8.195$. This means that for the corresponding
values of $\gamma$ we can lower $\omega_1=\omega_2$ down to zero.  
For $\omega_1=\omega_2=0$, the effective potential
\begin{equation}
 V_{eff}=\frac{1}{2}\left(\omega_1^2 \phi_1^2 + \omega_2^2 \phi_2^2\right) - V(\phi_1,\phi_2)
\end{equation}
is nowhere positive and the solutions are unphysical. Nevertheless, let us stress that we can have solutions
with arbitrary small charges (and frequencies). Note that this feature is not a consequence of the
interaction but is also present in the single $Q$-ball case for an appropriate choice of $\beta$ different
from that used in \cite{vw,kk1,kk2}.

It is apparent in
Figs.\ref{alp0q} and \ref{alp0m}, where we show the charges $Q_1=Q_2$ and the mass $M$ of
the  $Q$-ball solutions for negative values of $\gamma$, that solutions exist
down to $\omega_1=\omega_2=0$. The limiting solutions have vanishing charges, but non-vanishing
mass.
 
\begin{figure}[!htb]
\centering
\leavevmode\epsfxsize=15.0cm
\epsfbox{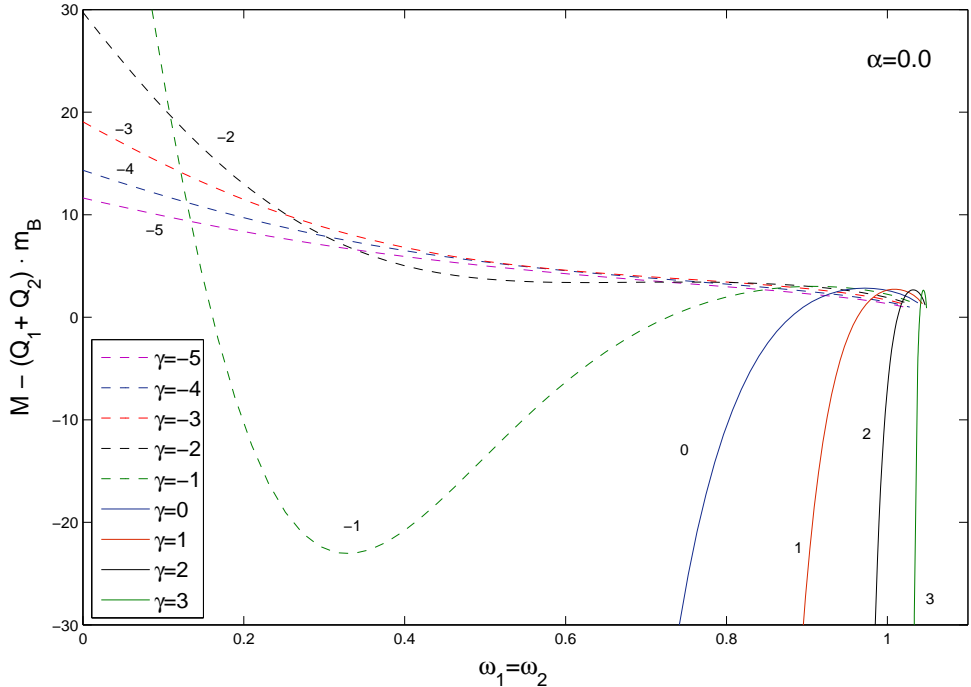}\\
\caption{\label{alp0binding} The dependence of the binding energy $M-(Q_1+Q_2)m_B$ of two interacting $Q$-balls on $\omega_1=\omega_2$ is shown for  different values of $\gamma$. The small numbers indicate the value of $\gamma$. }
\end{figure}

We also show the binding energy $M-(Q_1 m_{B,1} + Q_2 m_{B,2})=M-(Q_1 + Q_2 )m_B$ of the solutions in Fig.\ref{alp0binding}.
This quantity compares the mass of the interacting $Q$-balls $M$ with the mass
of $Q_1+Q_2$ bosons with respective masses $m_{B,1}\equiv m_{B}$ and $m_{B,2}\equiv m_B$ and indicates
whether the $Q$-balls are stable. For negative and positive binding energy, we expect the solutions to be stable and unstable, respectively.
We observe that for $\gamma \geq 0$, the solutions are stable for nearly all values of the frequencies $\omega_1=\omega_2$, apart from values close to the maximal frequency.

This changes for negative values of $\gamma$. For $\gamma \leq -2$, the binding energy is positive for all values of $\omega_1=\omega_2$ indicating an instability of the solution.
For $\gamma=-1$, the solutions are stable in the interval $\omega_1=\omega_2 \in [0.15 : 0.7]$, but unstable for all other values of the frequency.

\begin{figure}[!htb]
\centering
\leavevmode\epsfxsize=15.0cm
\epsfbox{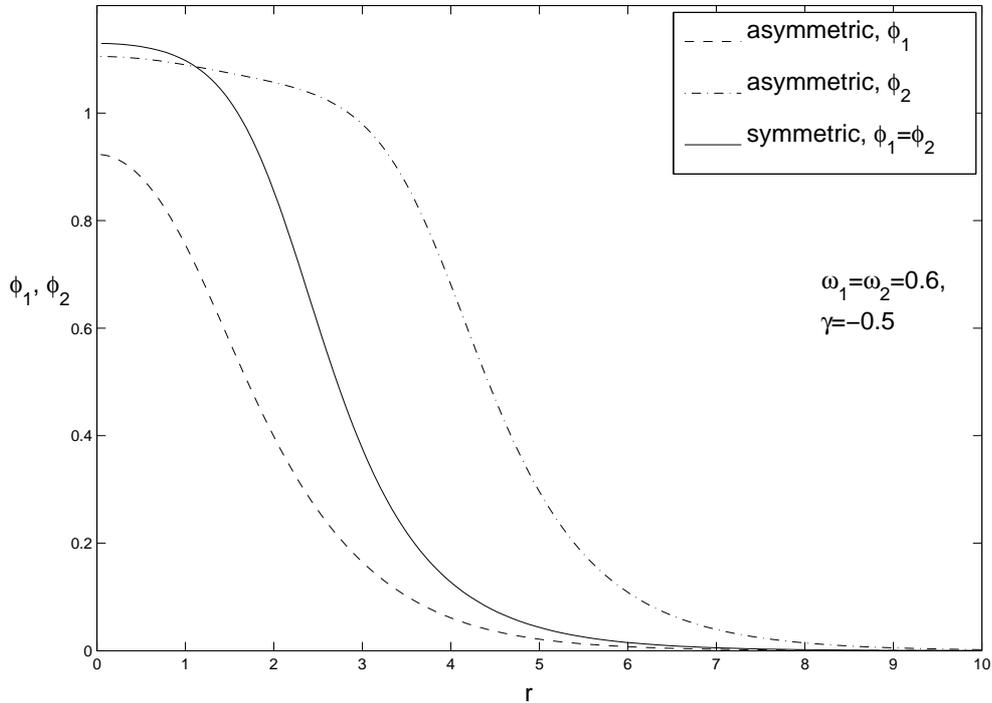}\\
\caption{\label{sas1} The profiles of the functions $\phi_1$ and $\phi_2$ are shown 
for symmetric $Q$-ball solutions and asymmetric $Q$-ball solutions, respectively. Here $\omega_1=\omega_2=0.6$ and $\gamma=-0.5$.  }
\end{figure}

For sufficiently strong interaction between the $Q$-balls and $\omega_1=\omega_2$, we observe a new phenomenon.
For  fixed values of the interaction parameter $\gamma$ and the frequency $\omega_1=\omega_2$, we find that
two types of solution exist. One solution has $\phi_1=\phi_2$ (``the symmetric solution'' in the following, see discussion above), while the second solution has $\phi_1\neq \phi_2$ (``the asymmetric solution'' in the following). This is demonstrated in Fig.\ref{sas1}
for $\gamma=-0.5$ and $\omega_1=\omega_2=0.6$ where we give the profiles of $\phi_1$ and $\phi_2$.
For the symmetric case, $\phi_1=\phi_2$, while for the asymmetric case $\phi_1\neq \phi_2$.
Note that in this case, we present a solution for which $\phi_1(0) < \phi_2(0)$, but that due to the symmetry
of the equations a second solution with $\phi_1$ and $\phi_2$ interchanged exists. In this sense, there is -- in fact -- not only one asymmetric solution for a fixed $\omega_1=\omega_2$, but two.

 Moreover,
it is obvious that $\phi_1/\phi_2$ is not simply a constant. We observe that the asymmetric solutions
have much higher energy than the symmetric ones, e.g. for the case plotted in Fig.\ref{sas1}, we have
$E_{symmetric}\approx 146$ for the energy of the symmetric solution, while the energy of the asymmetric
solution is $E_{asymmetric}\approx 288$.
For fixed $\gamma$, we observe that the asymmetric solutions exist only up to a maximal value of $\omega_1=\omega_2\equiv \omega$, which we
call $\omega_{cr}$ in the following. At $\omega_{cr}$, the asymmetric solutions join the branch
of symmetric solutions. We find that $\omega_{cr}$ depends crucially on the choice of $\gamma$. For
$\gamma \geq -0.195$, asymmetric solutions do not exist at all, while the critical value of $\omega$ increases
with decreasing (and negative) $\gamma$, e.g. we find $\omega_{cr}(\gamma=-0.5)\approx 0.74$
while $\omega_{cr}(\gamma=-1.0)\approx 0.82$.
This means that the stronger the repulsion between the $Q$-balls, the higher the frequency of the asymmetric
solution can be.

\begin{figure}[!htb]
\centering
\leavevmode\epsfxsize=16.0cm
\epsfbox{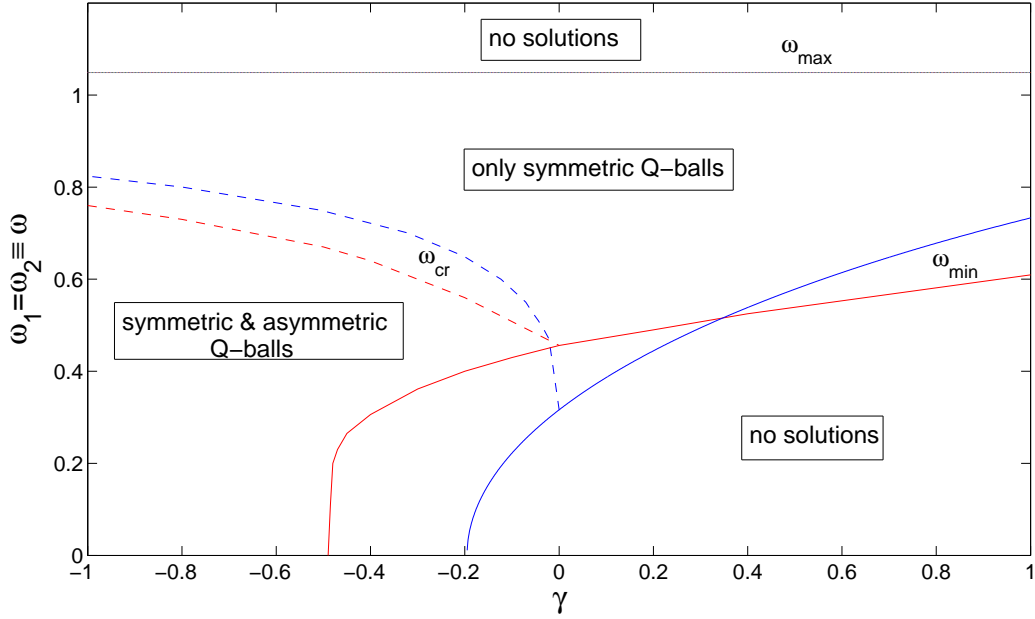}\\
\caption{\label{domain_alp0} The domain of existence for $Q$-balls with $\omega_1=\omega_2\equiv \omega$ is shown in the $\omega$-$\gamma$-plane for $\alpha=0$ (blue) and $\alpha=0.1$ (red), respectively.}
\end{figure}

The domain of existence of $Q$-ball solutions with $\omega_1=\omega_2\equiv \omega$ is shown
in Fig.\ref{domain_alp0} (blue curves). The curves for $\omega_{min}$ and $\omega_{max}$ are given by (\ref{omegamax}) and (\ref{omegamin}), while the curve $\omega_{cr}$ has been determined numerically.
As is obvious from this plot, asymmetric solutions exist only for $\gamma \leq -0.195$.

Let us emphasize that whenever asymmetric solutions exist, a corresponding symmetric solution
with lower energy is also present. In a concrete physical setting, we would thus expect the asymmetric
solutions to be unstable with respect to a decay into the symmetric solutions.

\subsubsection{$\omega_1\neq \omega_2$}

\begin{figure}[!htb]
\centering
\leavevmode\epsfxsize=13.0cm
\epsfbox{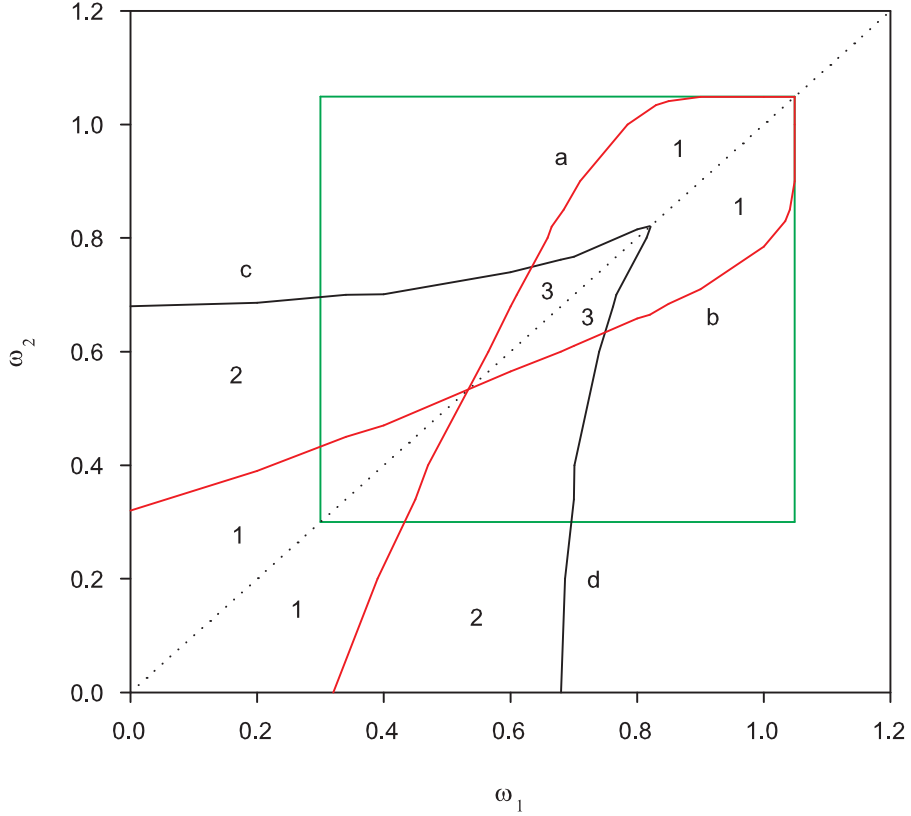}\\
\caption{\label{domain_0000_b} The domain of existence of non-rotating
Q-balls is shown in the $\omega_1$-$\omega_2$-plane for a repulsive interaction with $\gamma=-1$.}
\end{figure}

\begin{figure}[!htb]
\centering
\leavevmode\epsfxsize=15.0cm
\epsfbox{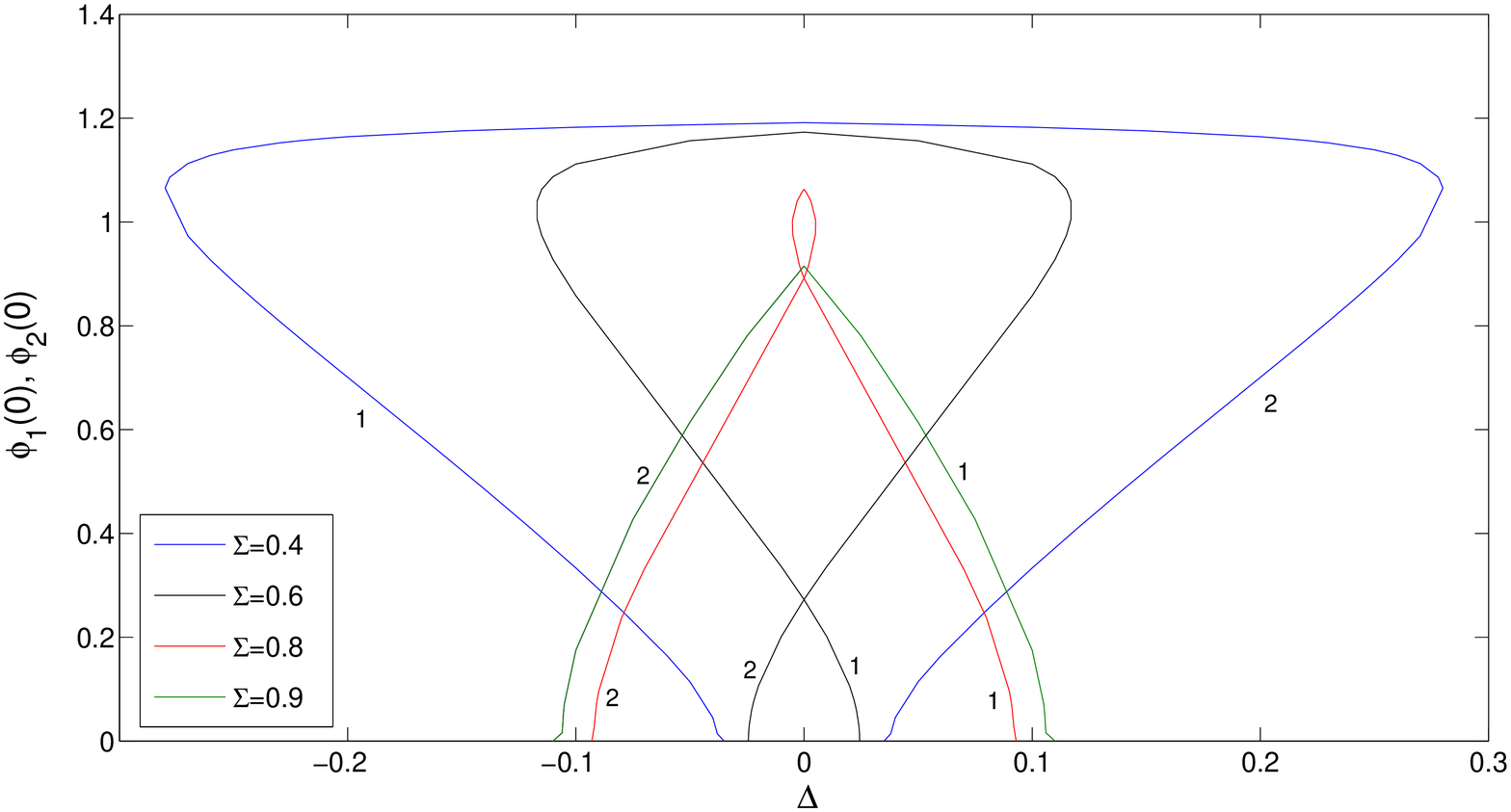}\\
\caption{\label{pattern_gamma_m1} The value of $\phi_1(0)$, respectively $\phi_2(0)$
is shown in dependence on $\Delta$ for different values of $\Sigma$. The labels ``1'' and ``2''
refer to the components $\phi_1$ and $\phi_2$.}
\end{figure}

\begin{figure}[!htb]
\centering
\leavevmode\epsfxsize=15.0cm
\epsfbox{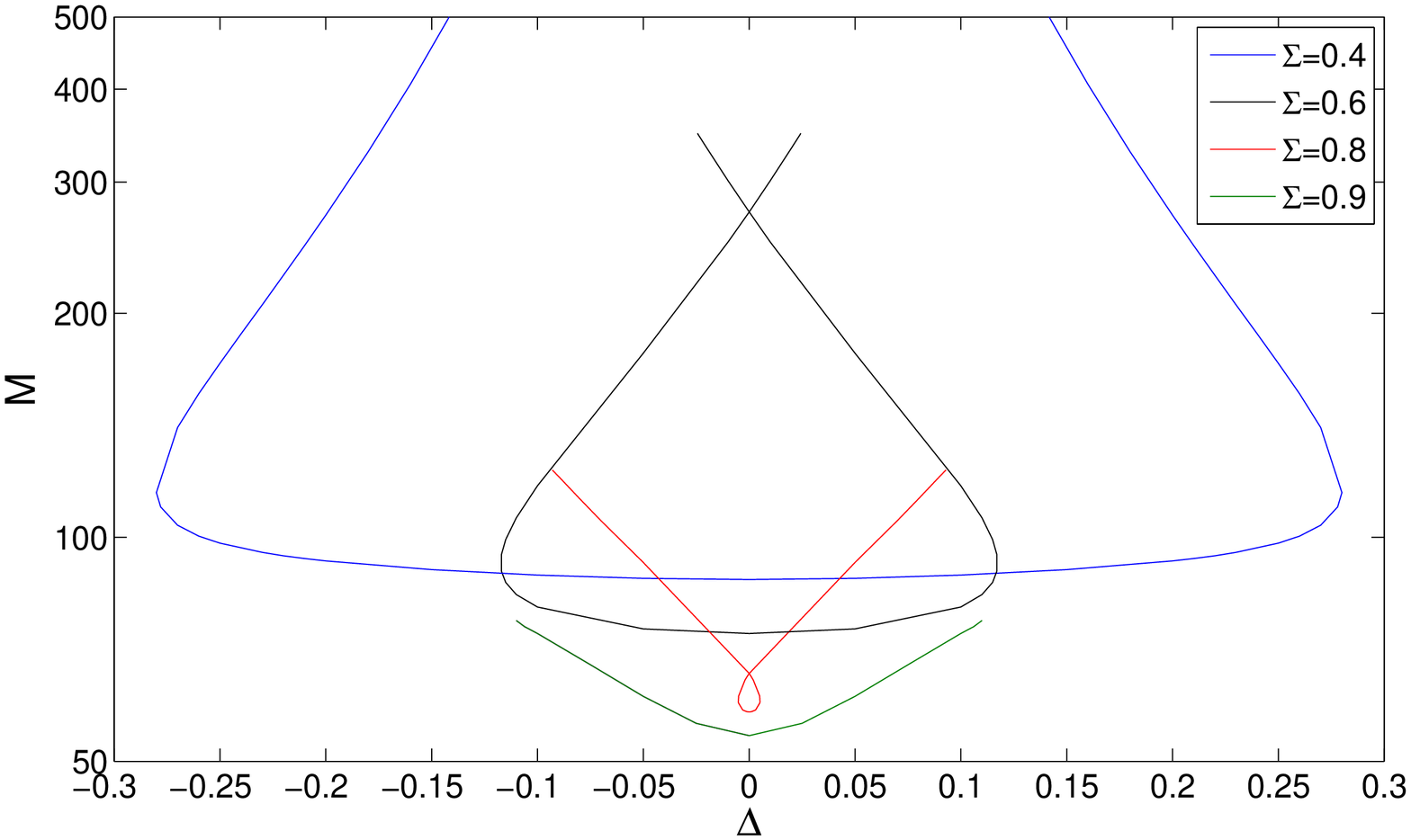}\\
\caption{\label{mass_gamma_m1} The value of the mass $M$
is shown in dependence on $\Delta$ for different values of $\Sigma$. }
\end{figure}

In this subsection, we would like to discuss the pattern of solutions
describing the repulsive interaction of two $Q$-balls with non-equal frequencies.
In the case $\gamma = 0$, the solutions exist in a rectangle of the $\omega_1$-$\omega_2$-plane which
consists of the direct product $[\omega_{1,min},\omega_{1,max}]\times [\omega_{2,min},\omega_{2,max}]$.
This is represented by the green rectangle
in Fig. \ref{domain_0000_b}. For $\gamma < 0$ (we set $\gamma = -1$)
the pattern of solutions changes non-trivially. Since the parameters of the 
potential are chosen in such a way that the equations are symmetric under
$\phi_1 \leftrightarrow \phi_2$, the domain should be symmetric under
the reflection $\omega_1 \leftrightarrow \omega_2$ . Accordingly the discussion of the domain is
easier by using the parameters $\Sigma \equiv (\omega_1 +\omega_2)/2$ and $\Delta \equiv (\omega_2-\omega_1)/2$.
We observe that up to three branches of solutions exist depending on the choice of $\Sigma$ and $\Delta$. It turns out that these various branches interconnect the symmetric and asymmetric solutions present in the case $\omega_1=\omega_2$. This is demonstrated in Fig.\ref{domain_0000_b}, where we give the domain of existence
of two $Q$-balls for $\gamma=-1$ in the $\omega_1$-$\omega_2$-plane.  
The labels $1$, $2$, $3$ in the figure refer to the number of solutions, while the labels $a$, $b$, $c$, $d$ are discussed
in the following. 
It turns out that the domain of existence can be separated into four subdomains.
To illustrate this, we plot $\phi_1(0)$ and $\phi_2(0)$ (see Fig.\ref{pattern_gamma_m1}) as well as the mass $M$ of the solutions (see Fig.\ref{mass_gamma_m1}) for four different fixed values of $\Sigma$ in dependence on $\Delta$. Note that the domain of existence is given in the $\omega_1$-$\omega_2$-plane,
but that we discuss the solutions using $\Delta$ and $\Sigma$. Constant values of $\Sigma$ correspond
to diagonals that fulfill $\omega_2= 2\Sigma-\omega_1$ in the $\omega_1$-$\omega_2$-plane.
These diagonals can be easily followed in Fig.\ref{domain_0000_b}.

\begin{enumerate}
\item For $0.96 < \Sigma < \omega_{max}$ the symmetric solution $\phi_1=\phi_2$ gets  deformed and exists for
 $-\Sigma/2 + \omega_{max} > \Delta > \Sigma/2 - \omega_{max} $. On the boundaries of this
interval, the charges and mass of the solutions diverge (like in the non-interacting case).
\item For $0.82 < \Sigma < 0.96$ (see $\Sigma=0.9$ in Fig.s \ref{pattern_gamma_m1}, \ref{mass_gamma_m1})  interacting solutions exist for $\Delta_b < \Delta \ < \Delta_a$ (with $\Delta_b = - \Delta_a$).
At $\Delta =  \Delta_{a,b}$ 
one of the two scalar fields vanishes identically and the limiting solution
does no longer represent interacting $Q$-balls, but only single $Q$-balls. The values of $\Delta_{a,b}$ 
are represented by the red lines  labelled $a$ and $b$ respectively.   
\item For $0.53 < \Sigma < 0.82$ (see $\Sigma=0.8$ and $\Sigma=0.6$ in Fig.s \ref{pattern_gamma_m1}, \ref{mass_gamma_m1}), a new phenomenon occurs. When deforming the solution $\phi_1=\phi_2$
by means of  $\Delta > 0$, we produce a branch of solutions which terminates at some
critical value of $\Delta$, say $\Delta = \Delta_c > 0$. This new critical value is
 represented by the black line and by the label ``c''.
A second branch of solutions then exist on $\Delta \in [\Delta_c,\Delta_b]$. On this second branch,
the mass of the solutions is higher than on the main branch. The second branch
extends in particular into the region $\omega_1 = \omega_2$ but the corresponding
solution has $\phi_2 \neq \phi_1$. This branch of solutions clearly connects the symmetric
with the asymmetric solution for $\omega_1=\omega_2$. 
Deformation of the solution $\phi_1=\phi_2$
by means of  $\Delta < 0$ is also possible and leads  
 of course to the occurrence of a third set of solutions for $\Delta \in [\Delta_a,\Delta_d]$.
\item Finally, for $\Sigma < 0.53$ (see $\Sigma=0.4$ in Fig.s \ref{pattern_gamma_m1}, \ref{mass_gamma_m1}) the symmetric solutions at $\Delta =0$ get deformed for $\Delta > 0$ up to a maximal value $\Delta_c$ (limiting
curve labelled ``c'' in Fig.\ref{domain_0000_b}). Starting from this value of $\Delta$ a second branch of
solutions exists for $\Delta \in [\Delta_c,\Delta_a]$, where $\Delta_a$ denotes the value of
$\Delta$ on the curves  labelled ``a''. For $\Delta < 0$, the pattern is similar, but the deformed symmetric
solutions exist down to $\Delta_d$ and from there a second branch of solutions exist up to $\Delta_b$.

\end{enumerate}

\subsection{ $\alpha \neq 0$~: Boson stars}
For $\alpha\neq 0$ the solutions are self-gravitating $Q$-balls, so-called boson stars.
In this case the solutions interact both via the potential interaction as well as through
gravity. 

\begin{figure}[!htb]
\centering
\leavevmode\epsfxsize=15.0cm
\epsfbox{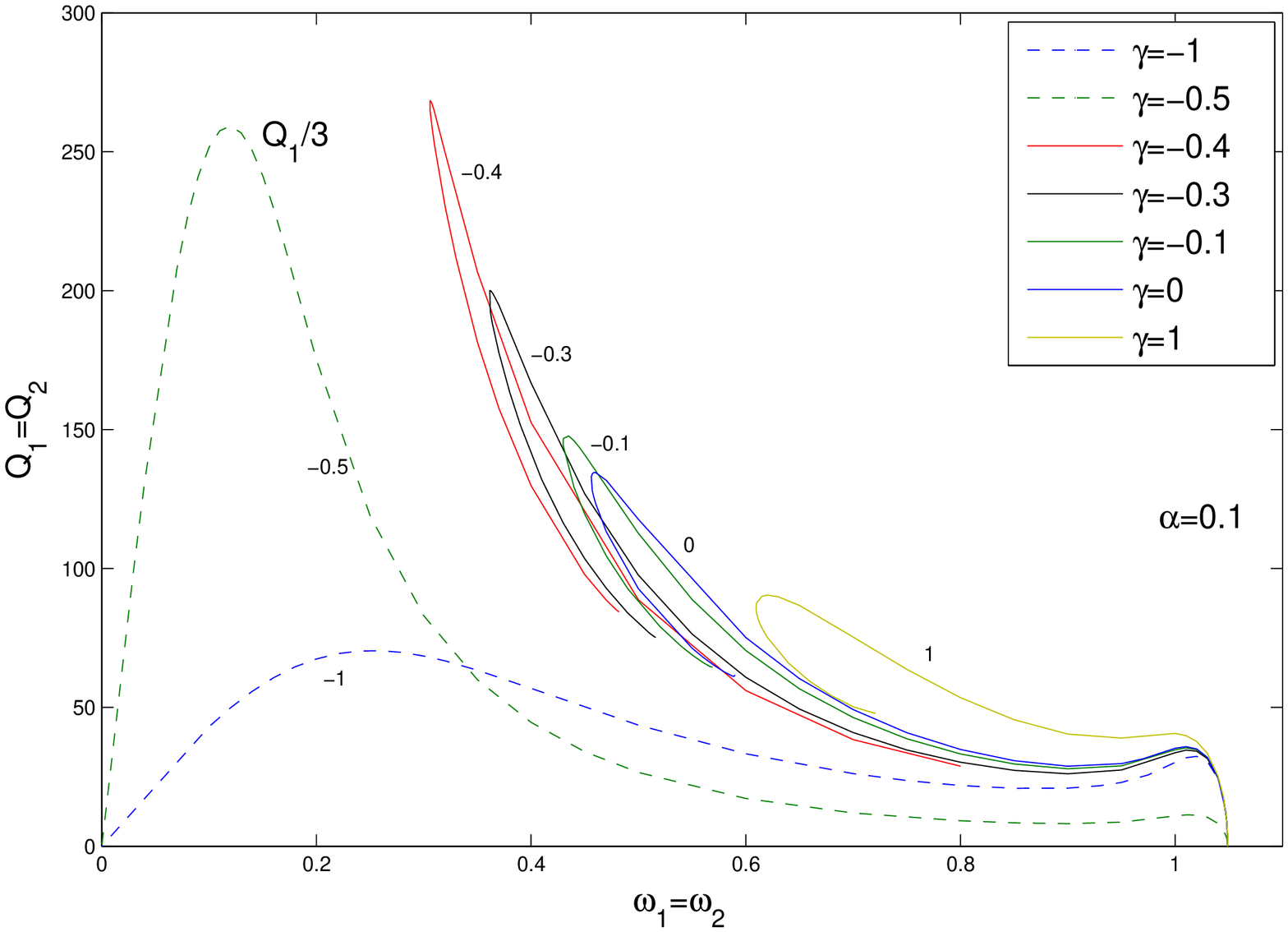}\\
\caption{\label{alp01q} The dependence of the charges $Q_1=Q_2$ of two interacting boson stars  on $\omega_1=\omega_2$ is shown for  different values of $\gamma$ and $\alpha=0.1$. The small
numbers indicate the value of $\gamma$. Note that we plot $Q_1/3=Q_2/3$ for $\gamma=-0.5$. }
\end{figure}

\begin{figure}[!htb]
\centering
\leavevmode\epsfxsize=15.0cm
\epsfbox{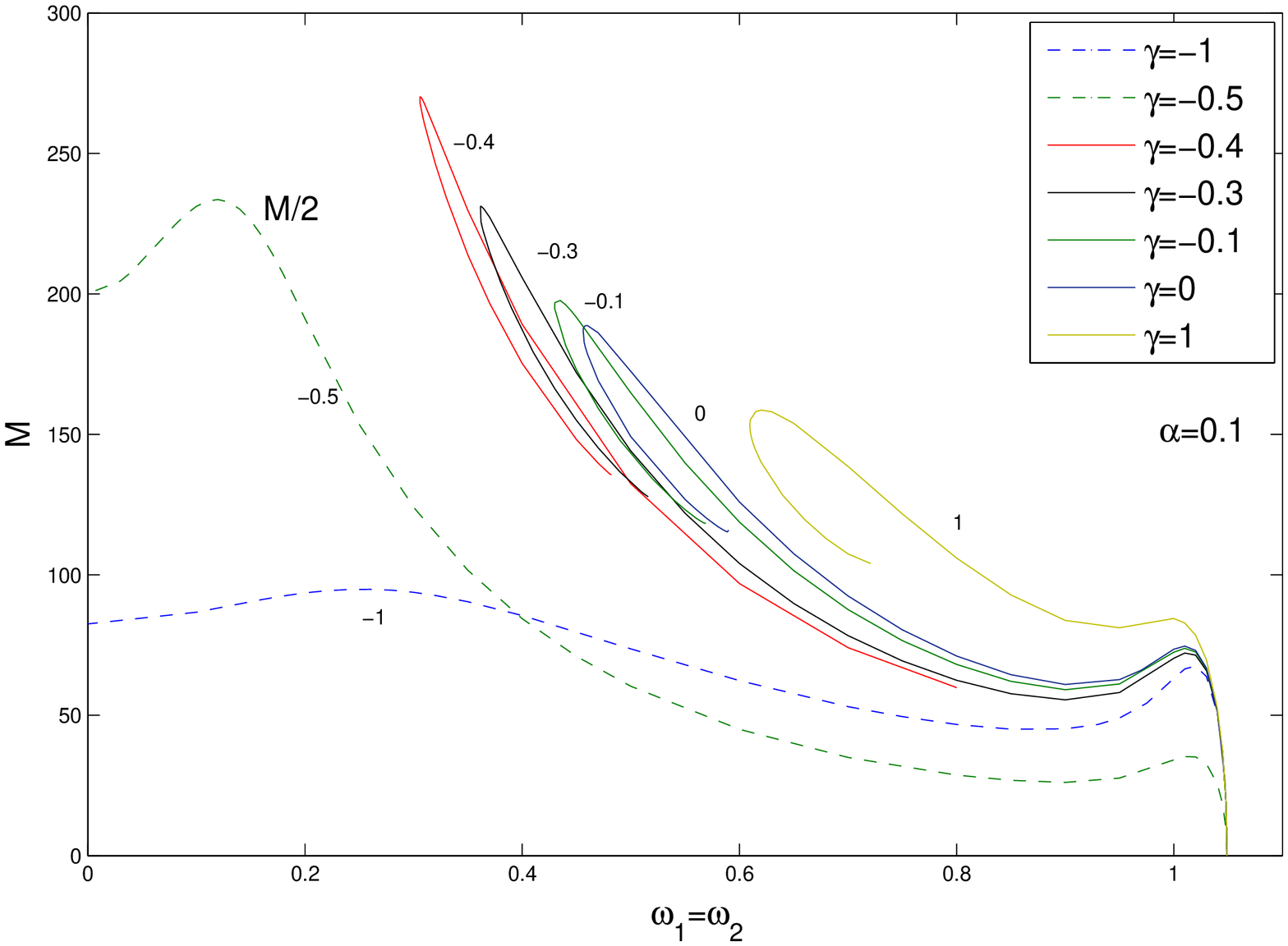}\\
\caption{\label{alp01m} The dependence of the mass $M$ of two interacting boson stars on $\omega_1=\omega_2$ is shown for  different values of $\gamma$ for $\alpha=0.1$. The small numbers
indicate the value of $\gamma$. Note that we plot $M/2$ for $\gamma=-0.5$.}
\end{figure}

\subsubsection{$\omega_1=\omega_2$}
\begin{figure}[!htb]
\centering
\leavevmode\epsfxsize=15.0cm
\epsfbox{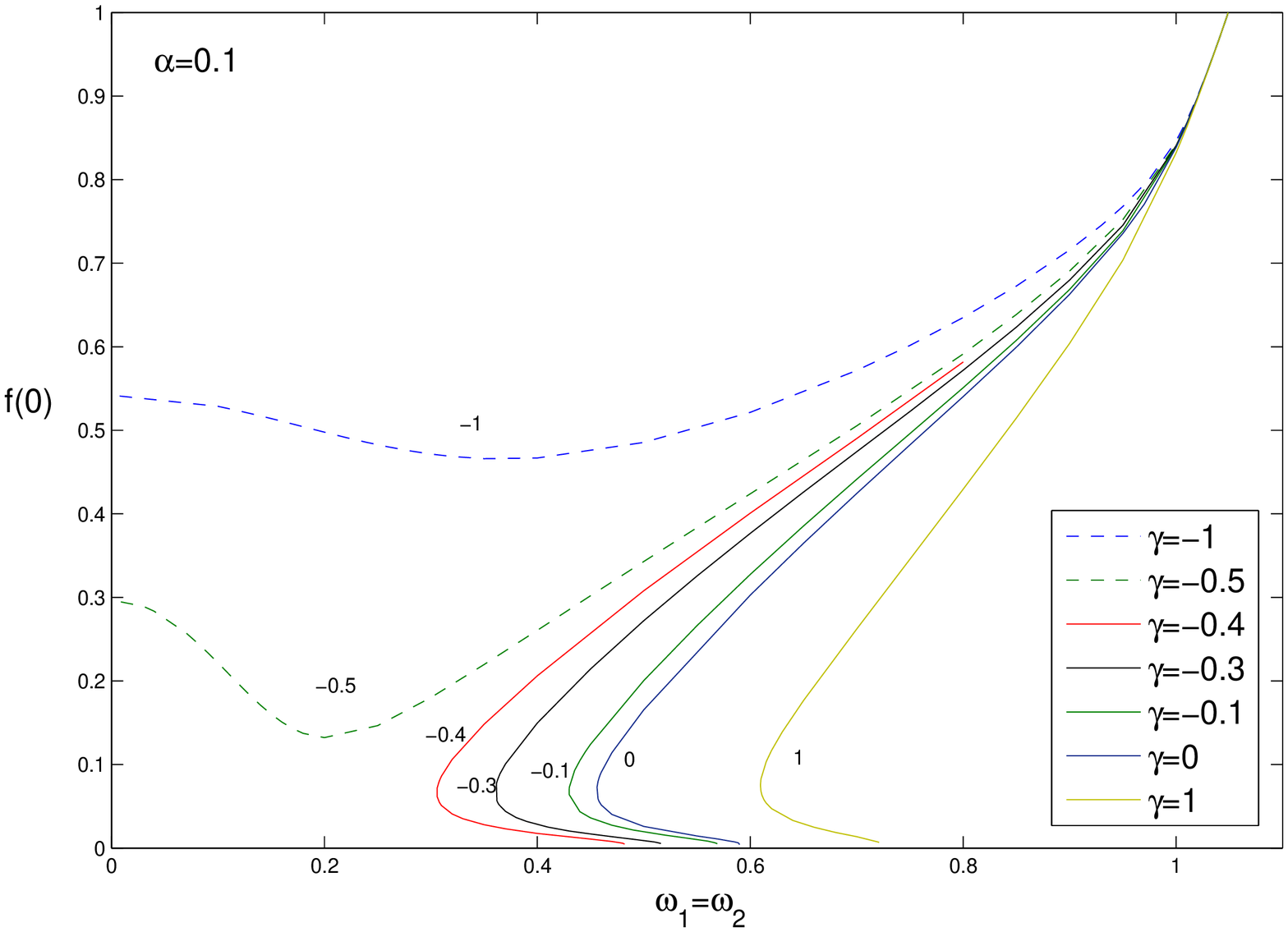}\\
\caption{\label{alp01f0} The dependence of the value of the metric function $f$ at the origin, $f(0)$ is shown for interacting boson stars for different values of $\gamma$ and $\alpha=0.1$. }
\end{figure}

First, we have fixed $\alpha=0.1$ and have studied the dependence of the charges and the mass
on the interaction parameter $\gamma$ for the symmetric case $\phi_1=\phi_2$ (for the case $\phi_1\neq \phi_2$ see the end of this section). Our results are shown in Figs.\ref{alp01q}, \ref{alp01m} and \ref{alp01f0}. Like in the flat space-time limit, the solutions
exist on a finite domain of the frequency.
As for single boson stars \cite{kk1, kk2}, the charges $Q_1=Q_2$ and the mass $M$ tend to zero for
$\omega_1 \rightarrow \omega_{1,max}$ ($\omega_2 \rightarrow \omega_{2,max}$).
The maximal value of $\omega_1=\omega_2$ seems to be practically independent of $\alpha$ and 
is equal to the flat space-time value. This is shown in Figs.\ref{alp01q} and \ref{alp01m}, where we give the dependence of the charges $Q_1=Q_2$ and the mass $M$, respectively, on the
frequency $\omega_1=\omega_2$. In this limit the value of the metric function $f(r)$ at the origin, $f(0)$, tends to one and since $f(\infty)=1$, we find $f(r)\equiv 1$.
This is plausible since in this limit, the mass tends to zero and there is thus no energy-momentum to curve the space-time.

For  $\omega_1=\omega_2$ tending to the minimal value we observe
an inspiralling of the charges as well as of the mass. This is very similar
to the case of single boson stars \cite{kk1,kk2}. At the same time, the value of
$f(0)$ tends to zero as can be seen in Fig.\ref{alp01f0}.
However, we find that this pattern of inspiralling exists only as long as the minimal value of the frequency is non-vanishing. When the solutions
exist down to $\omega_1=\omega_2=0$, the pattern changes. For $\gamma=-0.5$ and $\gamma=-1$ we observe 
that the charges $Q_1=Q_2$ tend to zero, but that the mass is non-vanishing.
The limiting solution is thus a non-trivial, static solution. Moreover, this solution
possesses non-trivial metric functions as is obvious from the fact that $f(0)\neq 1$ (see Fig.\ref{alp01f0}). This limiting solution is thus a self-gravitating static scalar field solution.
 
The value of $\gamma$ for which solutions exist down to $\omega_1=\omega_2=0$, $\gamma_0$, 
depends on $\alpha$. In the flat space-time limit, this value of $\gamma$ is given
analytically by (\ref{omegamin}) and we have $\gamma_0(\alpha=0)\approx -0.195$.
We find that this value of $\gamma_0$ decreases for increasing $\alpha$.
For $\alpha=0.1$, we have $\gamma_0(\alpha=0.1)\approx -0.49$. This is
shown in Fig.\ref{domain_alp0} (red curves). Interacting boson stars exist for all values of $\omega_1=\omega_2$ above the curves. 
Note that the $\alpha=0$ curve is the analytic curve given in  (\ref{omegamin}).

\begin{figure}[!htb]
\centering
\leavevmode\epsfxsize=15.0cm
\epsfbox{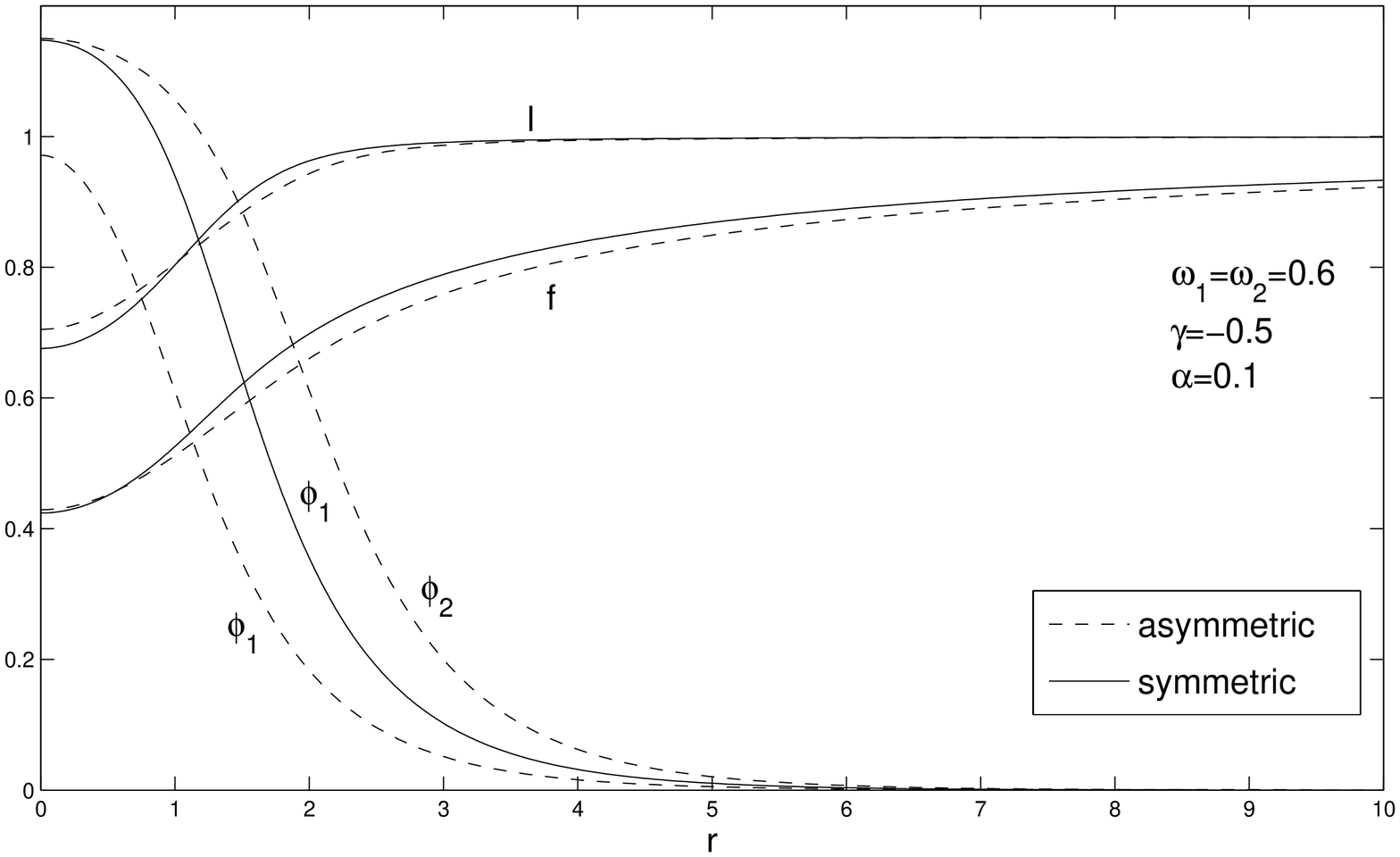}\\
\caption{\label{new_sas} The profiles of the scalar field functions $\phi_1$, $\phi_2$ and
of the metric functions $f$ and $l$
 are shown 
for symmetric boson stars and asymmetric boson stars, respectively. Here $\omega_1=\omega_2=0.6$, $\gamma=-0.5$ and $\alpha=0.1$. }
\end{figure}

Again, we observe that next to symmetric solutions, asymmetric solutions exist.
A typical asymmetric solution is shown in Fig.\ref{new_sas} for $\alpha=0.1$, $\gamma=-0.5$
and $\omega_1=\omega_2=0.6$. For comparison, we also plot the corresponding symmetric solution.
The profiles of the scalar field functions $\phi_1$ and $\phi_2$ look very similar to
what has been observed in the flat space-time limit, however, the boson stars have smaller radii
as the corresponding $Q$-balls for the same values of the parameters. The metric functions
of the symmetric solution differ only little from those of the asymmetric solution. We observe
that both $f(0)$ and $l(0)$ are slightly higher for the asymmetric solution. That the flat space-time phenomenon
persists when gravity is added is not that surprising. However, we have also 
studied this phenomenon for vanishing potential interaction $\gamma=0$ and have chosen $\alpha=0.1$ letting the boson
stars with $\omega_1=\omega_2\equiv \omega$ interact only via gravity.
We find that symmetric solutions exist on the interval $[\omega^{S}_{min}:\omega^{S}_{max}]\approx [0.456:1.041]$,
while asymmetric solutions exist on the interval $[\omega^{AS}_{min}:\omega^{AS}_{max}]\approx [0.355:0.462]$.
Thus, there is a small interval in which both symmetric as well as asymmetric solutions exist.
In this interval the mass of the symmetric solution is much smaller than that of the asymmetric
one. We would thus expect the asymmetric solutions to be unstable to decay to the symmetric
ones. However, for $0.355 < \omega < 0.456$ only asymmetric solutions exist. This is a completely new phenomenon
as compared to the flat space-time limit.

When studying the binding energy of the solutions, we observe that the curves
look qualitatively similar to the ones shown in Fig.\ref{alp0binding}, this is why we don't present
an extra figure here. Quantitatively, boson stars are stronger bound than their
corresponding flat space-time counterparts, which is -- of course -- due to the attractive
nature of the gravitational interaction.


\subsubsection{$\omega_1\neq \omega_2$}
\begin{figure}[!htb]
\centering
\leavevmode\epsfxsize=15.0cm
\epsfbox{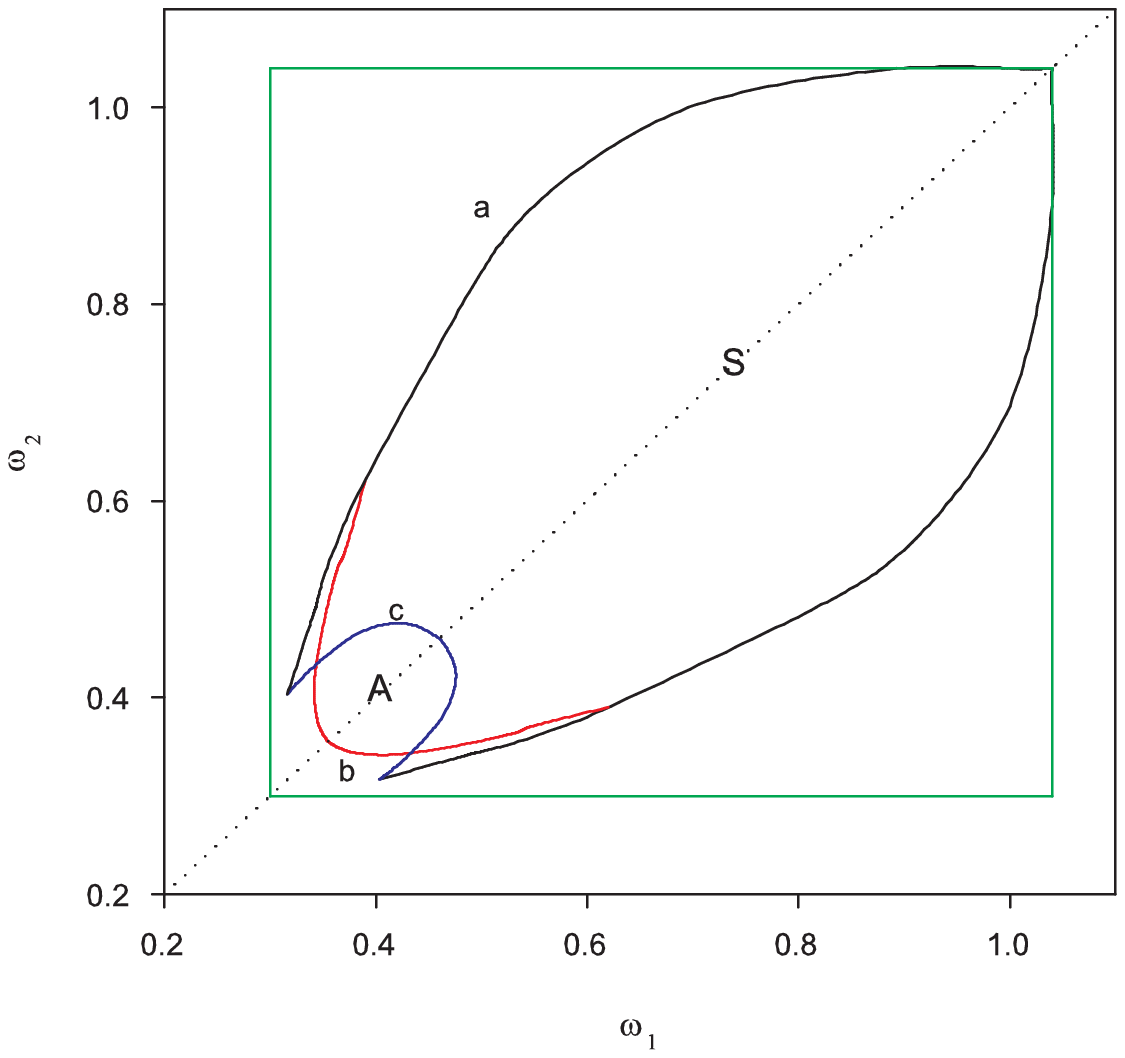}\\
\caption{\label{domain_alpha_01} The domain of existence of interacting
boson stars in shown in the $\omega_1$-$\omega_2$-plane for $\alpha=0.1$. }
\end{figure}

\begin{figure}[!htb]
\centering
\leavevmode\epsfxsize=15.0cm
\epsfbox{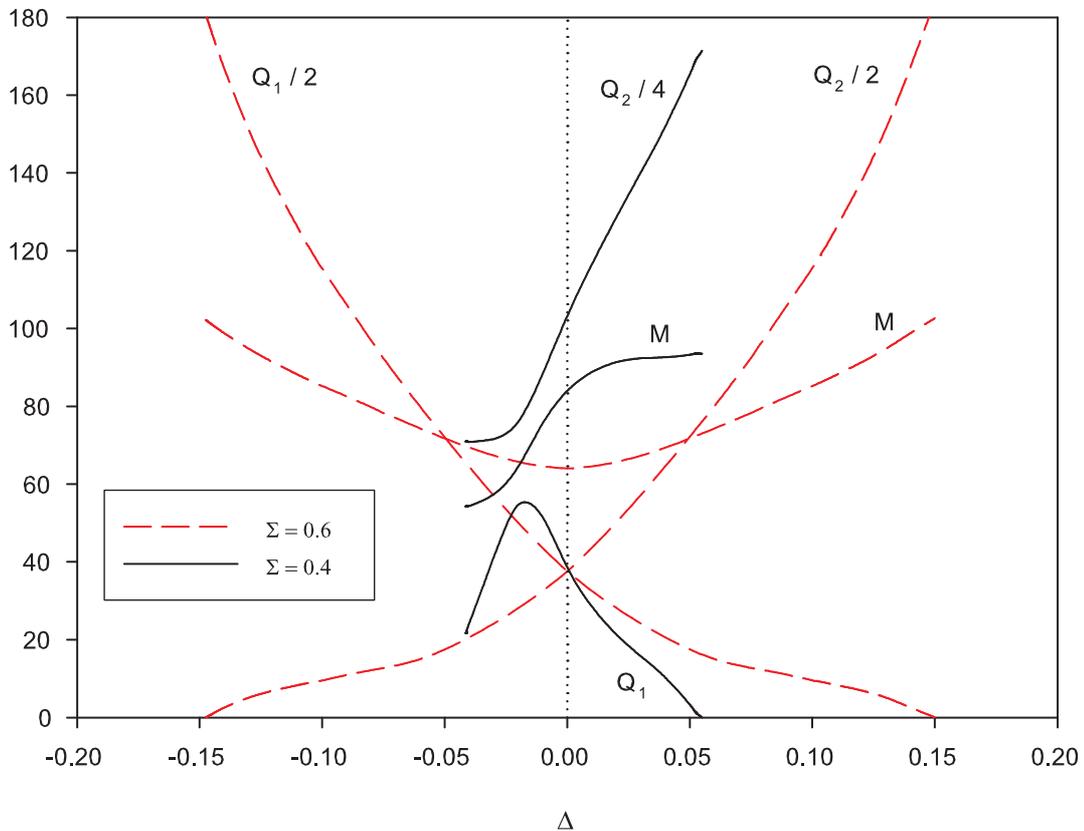}\\
\caption{\label{alpha_01_new} The value of the mass $M$ and the charges $Q_1$ and $Q_2$ are shown in dependence on $\Delta=(\omega_1-\omega_2)/2$
for two different choices of $\Sigma=(\omega_1+\omega_2)/2$. Here $\alpha=0.1$ and $\gamma=0$.}
\end{figure}

\begin{figure}[!htb]
\centering
\leavevmode\epsfxsize=15.0cm
\epsfbox{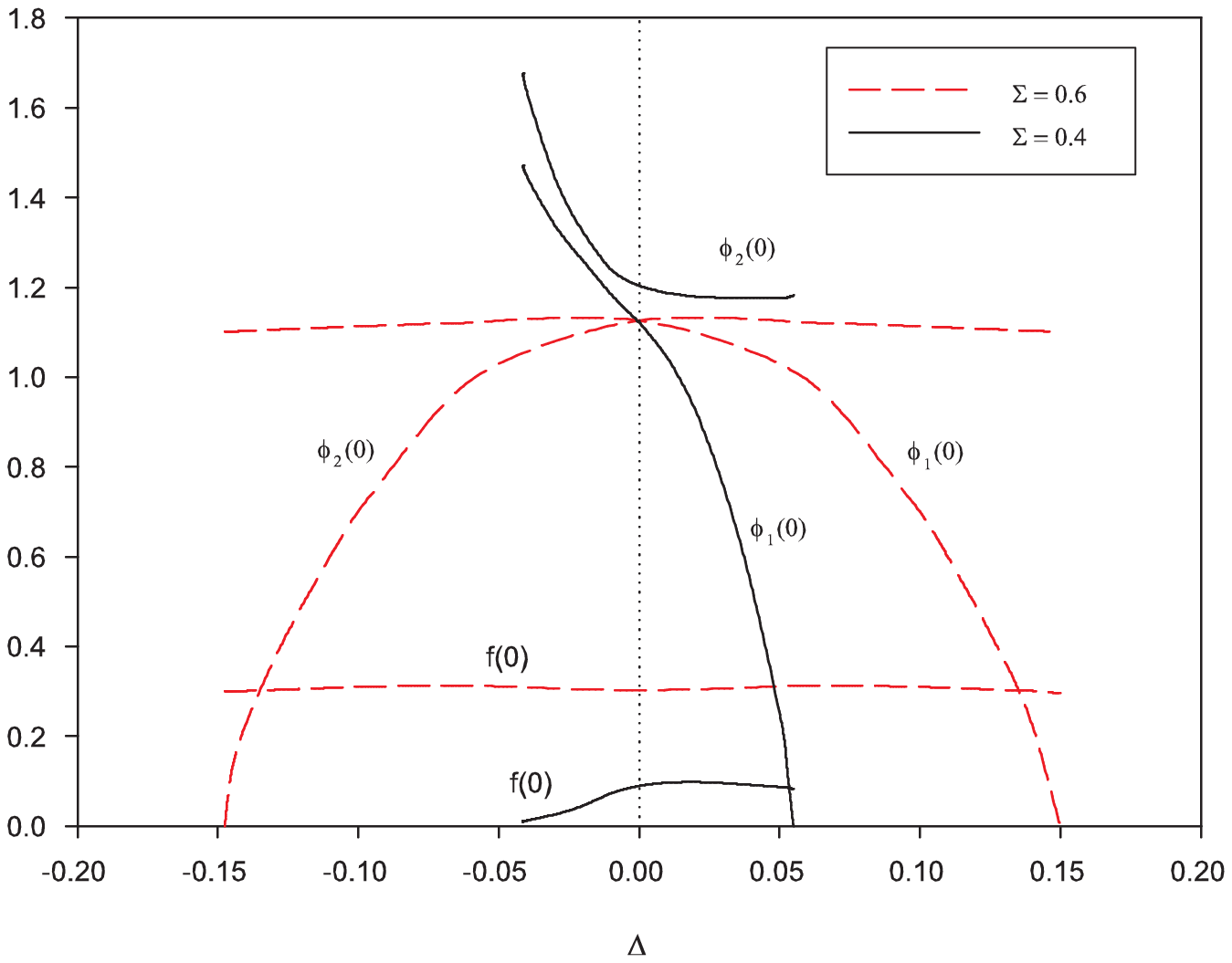}\\
\caption{\label{alpha_01} The value of the scalar field functions and of the metric function
$f$ at the origin $\phi_1(0)$, $\phi_2(0)$, $f(0)$ are shown in dependence on $\Delta=(\omega_1-\omega_2)/2$
for two different choices of $\Sigma=(\omega_1+\omega_2)/2$. Here $\alpha=0.1$ and $\gamma=0$.}
\end{figure}

Here, we concentrate on the influence of the gravitational interaction on the domain
of existence of non-rotating boson stars.
Let us recall that 
the solutions exist in a rectangle of the $\omega_1$-$\omega_2$-plane which
is described by the interval $[\omega_{1,min},\omega_{1,max}]\times [\omega_{2,min},\omega_{2,max}]$.
The domain of Q-balls in the $\alpha=0$ limit is shown in Fig. \ref{domain_alpha_01} (green lines).

For $\gamma=0$, $\alpha > 0$ (we have chosen $\alpha = 0.1$) the pattern changes
considerably. Again, it is useful to discuss the solutions by using $\Sigma=(\omega_1+\omega_2)/2$ and $\Delta=(\omega_2-\omega_1)/2$. Constant $\Sigma$s correspond again
to diagonals in the $\omega_1$-$\omega_2$-plane.
In the $\Delta=0$ limit,
the symmetric solutions $\phi_1=\phi_2$ exist for $\Sigma \in [0.456,1.041]$
and can de deformed into solutions with $\Delta \neq 0$. For $\Delta \to \Delta_a$ 
(resp. $\Delta \to -\Delta_a$)  the field $\phi_1$ (resp. $\phi_2$) vanishes identically and
the limiting solution represents a single boson star. This is demonstrated in Fig.s \ref{alpha_01_new},\ref{alpha_01} for $\Sigma=0.6$. In Fig.\ref{alpha_01_new}, it is apparent that for $\Delta \to \Delta_a\approx
0.55$, the charge $Q_1$ tends to zero (see also Fig.\ref{alpha_01}, where $\phi_1(0)\to 0$ in this
limit). For  $\Delta \to -\Delta_a\approx
-0.55$ we find $Q_2\to 0$. For $Q_i\to 0$, $i=1,2$, a single boson star remains that has finite mass and charge. 

The corresponding domain in the $\omega_1$-$\omega_2$-plane and the critical values $\pm \Delta_a$ are labelled by the symbol ``S'' and
the lines ``a'' in Fig. \ref{domain_alpha_01}, respectively. 

For $\Sigma \in [0.355, 0.465]$ and $\Delta = 0$, we observe that a pair of new solutions exists bifurcating
from the branch of symmetric solutions at $\Sigma \approx 0.465$. These solutions have $\phi_1 \neq \phi_2$
and correspond to asymmetric interacting boson stars. One of the solutions has $\phi_1(0) > \phi_2(0)$, while
the other has $\phi_1(0) < \phi_2(0)$. Note that the two solutions are, of course, related to each
other by the exchange $\phi_1 \leftrightarrow \phi_2$. 

Deforming the asymmetric solutions for $\Delta > 0$, it turns out that 
the solutions which have $\phi_1(0) <  \phi_2(0)$ at $\Delta =0$ will
have a vanishing scalar field $\phi_i(r)\equiv 0$ at $\Delta_b$ (denoted by the red line ``b'' in 
Fig.\ref{domain_alpha_01}). This is demonstrated in Fig.s \ref{alpha_01_new}, \ref{alpha_01} for $\Sigma=0.4$, where $Q_1=0$ and $\phi_1(0)=0$ (and hence also $\phi_1(r)\equiv 0$) at $\Delta_b \approx 0.06$, while
the mass $M$, the charge $Q_2$, $\phi_2(0)$ and $f(0)$ stay finite in this limit. On the other hand, if we deform
the solution which has $\phi_i(0) < \phi_j(0)$ at $\Delta=0$ for $\Delta < 0$, we find
that $f(0)$ tends to zero (after a possible inspiralling) at $\Delta_c$ (denoted by the blue line ``c'' in Fig.\ref{domain_alpha_01}). This   
phenomenon is demonstrated in Fig.\ref{alpha_01} for $\Sigma=0.4$ and $\phi_1(0) < \phi_2(0)$ at $\Delta=0$.
Clearly for $\Delta\approx -0.03$, we have $f(0)\approx 0$.

Note that if we would choose the asymmetric solution with $\phi_1(0) > \phi_2(0)$ at $\Delta=0$
the pattern would be the same, but with $\Delta_b \rightarrow -\Delta_b$ and $\Delta_c\rightarrow -\Delta_c$.

In the small domain between the curves ``b'' and ``c'' only solutions which are deformations
of the asymmetric solution at $\Delta=0$ (denoted
by ``A'') exist. This is completely different from the flat space-time case, where asymmetric
solutions exist only when also symmetric solutions are present for $\omega_1=\omega_2$.
This means that for appropriate choices of the frequencies, boson stars could exist
that have the same frequencies but different charges. These asymmetric solutions
would then be the generic solutions in this parameter range.
\section{Rotating solutions}
As pointed out already in previous sections, the equations for the fields $\phi_1$ and $\phi_2$
totally decouple in the limit $\gamma=0$, $\alpha=0$ and solutions of the different 
types available in the case of a single $Q$-ball can  arbitrarily be superposed. 
Solutions of the decoupled system can naturally be labelled
according to the ``quantum numbers'' characterizing each of the two scalar fields, say $k_i$, $l_i$ for $i=1,2$.
Any such configuration then gets deformed by the direct interaction if $\gamma\neq 0$  and/or 
by gravity if $\alpha\neq 0$.
The domain of existence in the $\omega_1$-$\omega_2$-plane is rectangular when $\alpha=\gamma=0$ and 
we observe this domain to be deformed when $\gamma$ and/or $\alpha$ are non-vanishing.
The study of the evolution of this domain turns out to be very involved. In this section,
we  present some of its  qualitative features for particular cases.
More precisely, we study rotating $Q$-balls (or boson stars) interacting with non-rotating $Q$-balls 
(or boson stars). We have chosen the fundamental non-rotating solution characterized by
$k_1=0$, $l_1=0$ to interact with the rotating solution with $k_2=1$, $l_2=1$.

We have solved the system of partial differential equations
(\ref{einstein}) and (\ref{KG}) subject to the appropriate boundary.
This has been done using the Partial differential equation solver FIDISOL
\cite{fidi}.
We have mapped the infinite interval of the $r$ coordinate $[0:\infty]$ to the
finite
compact interval $[0:1]$ using the new coordinate $z:=r/(r+1)$.  
We have typically used grid sizes of $150$ points in $r$-direction and $70$
points in $\theta$ direction. The solutions presented here have relative errors of $10^{-3}$
or smaller.

\begin{figure}[!htb]
\centering
\leavevmode\epsfxsize=15.0cm
\epsfbox{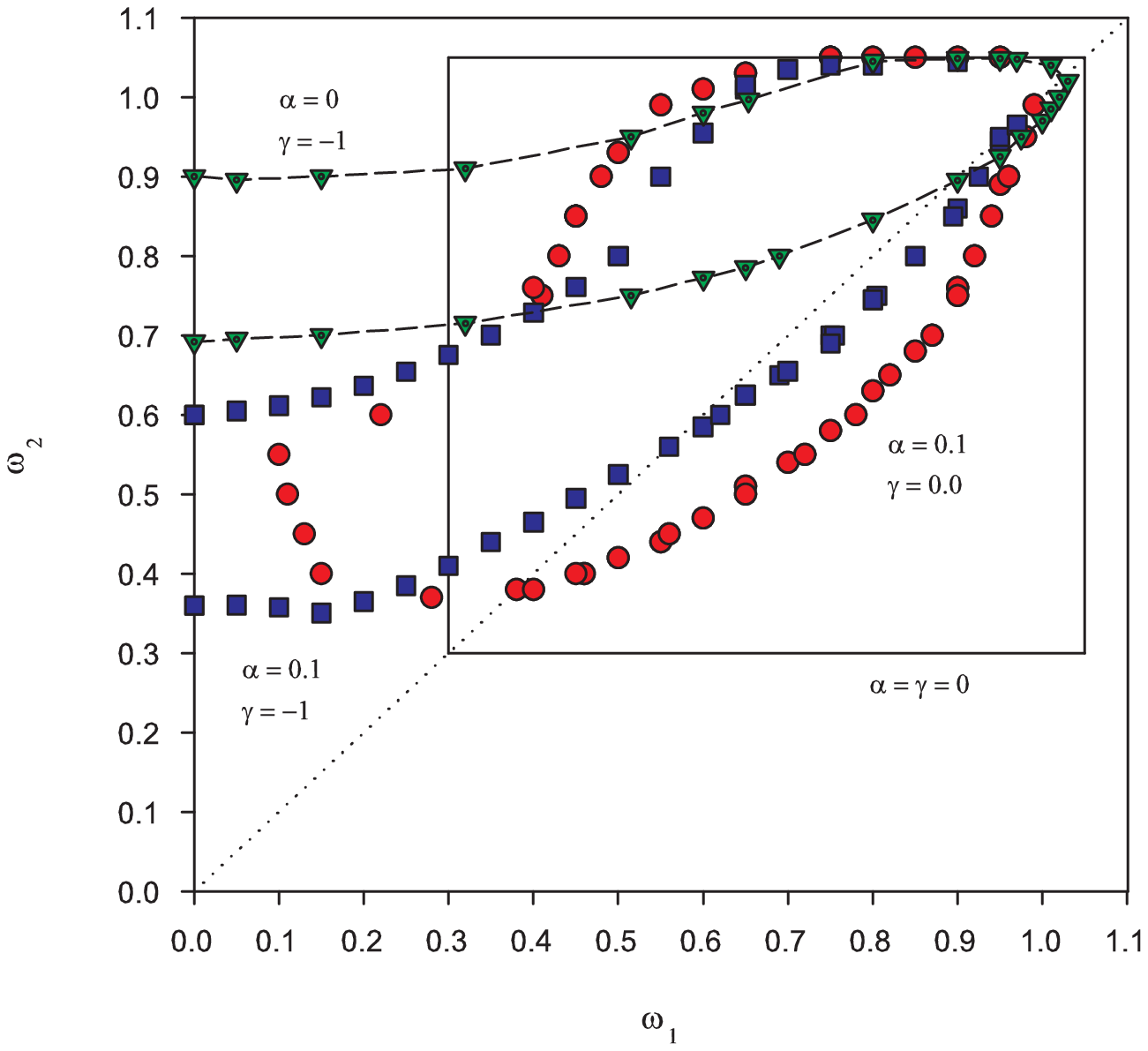}\\
\caption{\label{domain_00_11} 
The domains of existence of non-rotating $Q$-balls (boson stars) interacting with rotating
$Q$-balls (boson stars) are shown in the $\omega_1$-$\omega_2$-plane for different values of
$\gamma$ and $\alpha$.}
\end{figure}

\begin{figure}
\includegraphics[width=8cm]{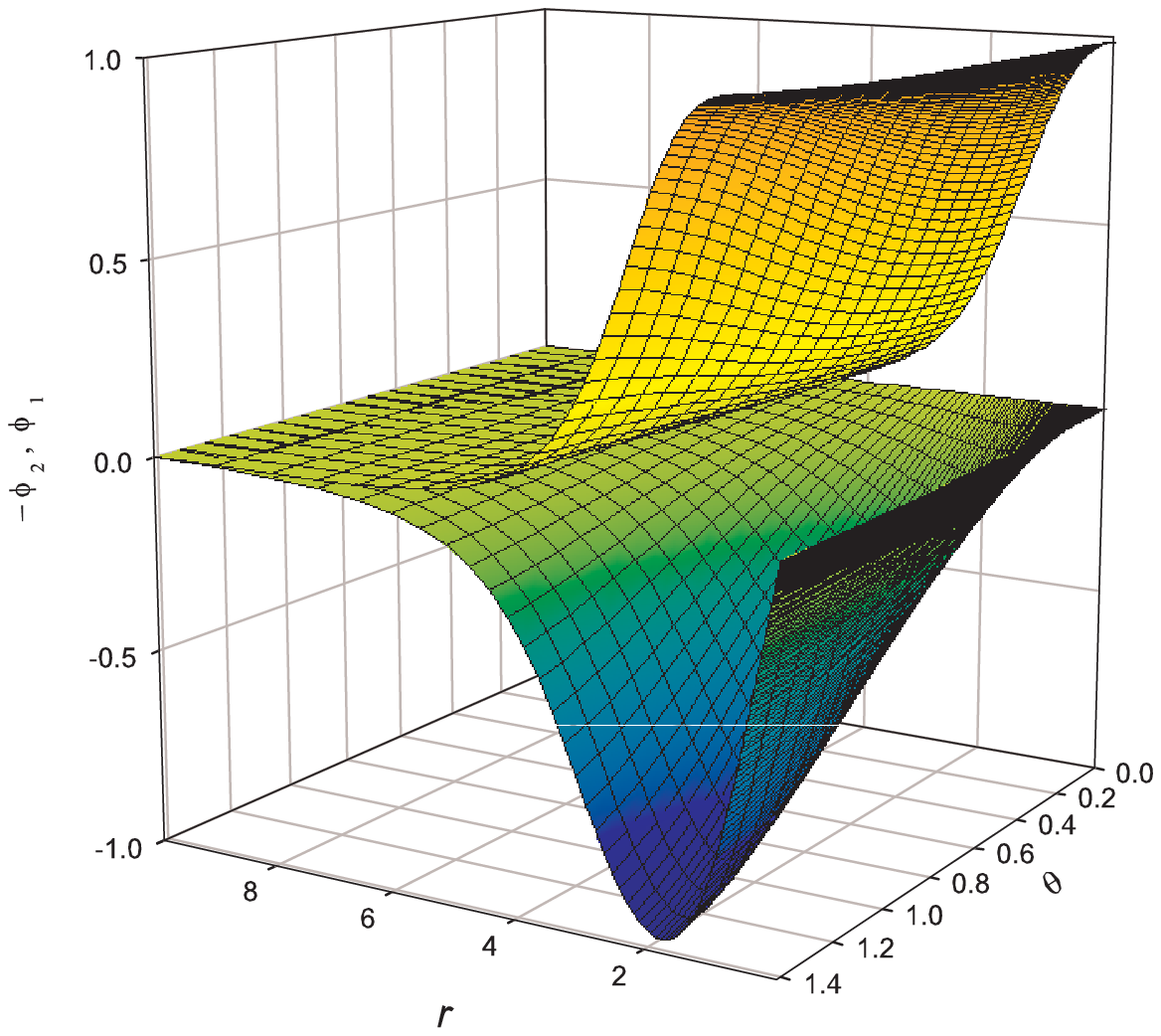} 
\includegraphics[width=8cm]{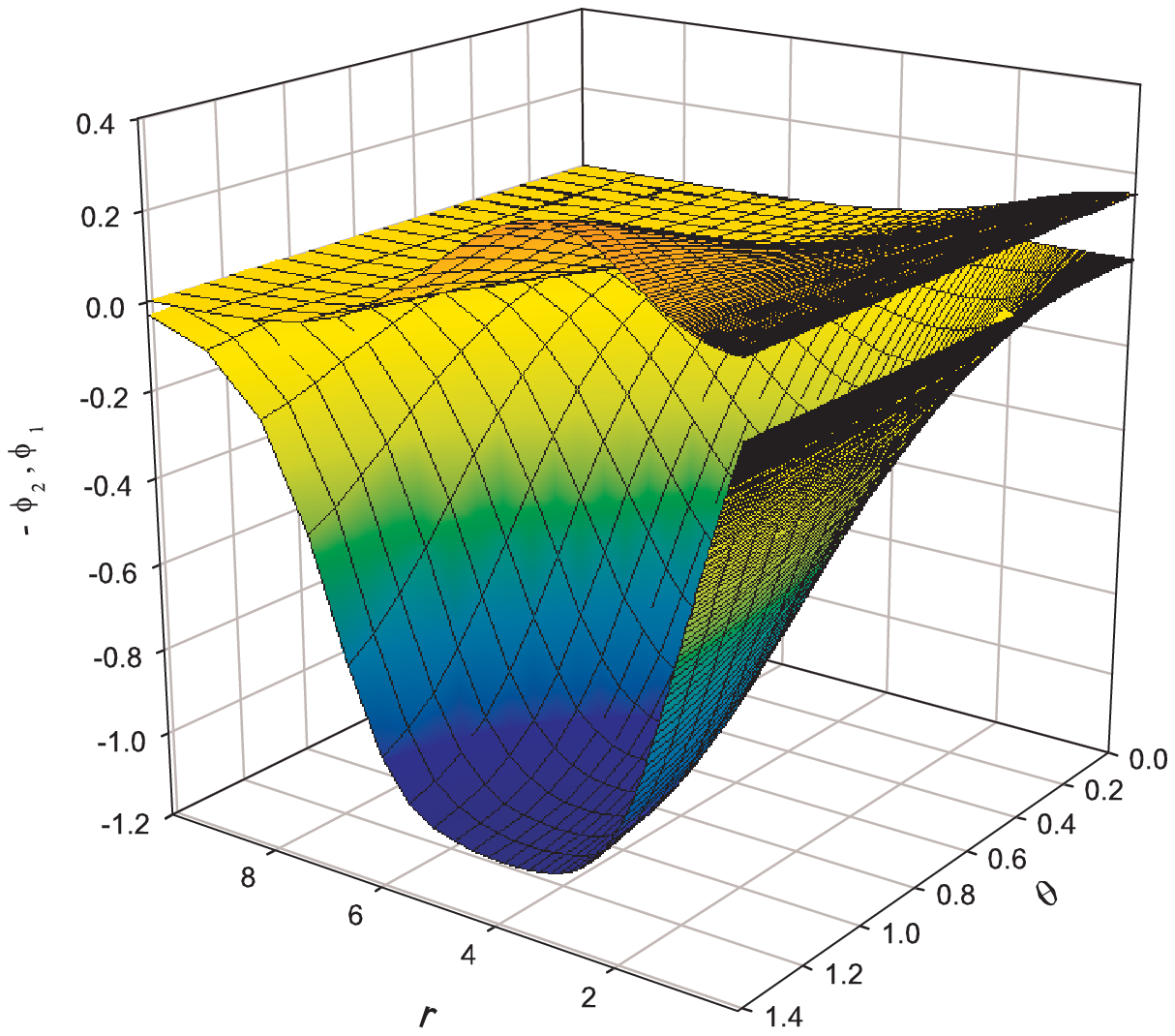}
\caption{ \label{fig_00_11} The scalar field functions $\phi_1$ and $\phi_2$ 
for a system of a non-rotating $Q$-ball with $k_1=l_1=0$ interacting with 
a rotating $Q$-ball with $k_2=l_2=1$ and $\gamma=-0.5$. The plot on the left and right
corresponds to  $\omega_1=0.5$, $\omega_2=0.85$ and 
      $\omega_1=0.5$, $\omega_2=0.525$, respectively.}
\end{figure}

\subsection{$\alpha=0$: Interacting $Q$-balls}
For $\gamma=0$, the domain of existence of the $k_1=0$, $l_1=0$ and $k_2=1$, $l_2=1$ solutions
consists of the rectangle
\be
           \omega_1 \in [\omega_{1,min}, \omega_{1,max}] \ \ , \ \ 
           \omega_2 \in [\omega_{2,min}, \omega_{2,max}]   \ .
\ee
This is indicated by the black line in Fig. \ref{domain_00_11}. 
The boundary of the rectangle is given by \cite{kk1}~:
\be
        \omega_{1,max} = \omega_{2,max} = \sqrt{1.1} \ \ , \ \ \omega_{1,min}=\sqrt{1/10}\approx \omega_{2,min} \ \  .
\ee
 In fact, on basis of numerical results, it has been conjectured in \cite{kk1} that $\omega_{2,min}=\omega_{1,min}$, but
 this equality could not be proved analytically. 
For both limits $\omega_i \to \omega_{i,max}$, $i=1,2$ or 
$\omega_i \to \omega_{i,min}$, $i=1,2$ the corresponding charges $Q_i$ diverge.
When $\omega_i \to \omega_{i,max}$, $i=1,2$, the fields $\phi_i$ spread completely over the
full interval of $r$, while for  $\omega_i \to \omega_{i,min}$, $i=1,2$, they become step-like.
 
We now discuss  qualitatively the deformation of the domain in  the case of interacting $Q$-balls, i.e. for $\gamma\neq 0$.
We have analyzed in detail the cases $\gamma = - 0.5$ and $\gamma = -1$
 which capture the main features. 
 The form of the domain is given in Fig. \ref{domain_00_11} 
 for $\gamma=-1$  by the line with green triangles.

\subsubsection{Fixed $\omega_1$}
For fixed $\omega_1$ the solutions exist on 
a finite interval of $\omega_2$. The limits of this interval depend only slightly on $\omega_1$, but strongly 
on $\gamma$.
E.g. for $\omega_1 = 0.5$ we find
$$
 \omega_2 \in [0.52,0.85] \ \ {\rm for} \ \ \gamma = -0.5 \ \ \ \ , \  \ \ \
 \omega_2 \in [0.75,0.95] \ \ {\rm for} \ \ \gamma = -1.0
 $$ 
 i.e. decreasing $\gamma$, the interval in $\omega_2$ for which solutions exist decreases
 and gets shifted  to higher $\omega_2$.
 For $\omega_2$ close to $\omega_{2,max}$, the profile of $\phi_1$ hardly deviates from the spherically symmetric solution
(i.e. $\phi_1$ has its maximum at the origin). 
The profiles of the two scalar functions are shown in Fig.\ref{fig_00_11} (left) for $\omega_1=0.5, \omega_2=0.85$ for $\gamma=-0.5$.
The solution stops for a reason which remains unclear, it is very likely
that a second branch of solution backbending from the first branch exists.
Should it exist, it would constitute
the counterpart of the branch of asymmetric solutions occurring  in the case of two interacting, non-rotating $Q$-balls (boson stars).

Decreasing $\omega_2$, the numerical results show that the function $\phi_1$ progressively
develops a local maximum on a ring in the equatorial plane. 
The function $\phi_1$ becomes progressively smaller,
while the ring where the function $\phi_2$  reaches its maximum
(typically $|\phi_2| \sim 1$)  becomes larger, spreading more and more over space.
This phenomenon is illustrated in Fig.\ref{fig_00_11} (right) for $\omega_1=0.5$, $\omega_2=0.525$ and $\gamma=-0.5$. The pattern is qualitatively
similar for other values of $\omega_1$, including $\omega_1=0$. 
Thus, the spherically symmetric, non-rotating $Q$-ball is progressively absorbed by the axially symmetric, rotating $Q$-ball  when the parameter $\omega_2$ decreases. 

\subsubsection{Fixed $\omega_2$}
Another striking feature of interacting $Q$-balls is that the direct interaction  allows for solutions with
arbitrarily small $\omega_1$ and hence charge $Q_1$. 
Fixing $\omega_2$ and varying
$\omega_1$ shows that interacting solutions exist for $\omega_1 > 0$ and that the 
function $\phi_1$ representing the non-rotating $Q$-ball spreads over the full $r$ interval when $\omega_1$ tends to $\omega_{1,max}$. Only the rotating component of the solution (i.e. $\phi_2$) survives in this limit.
For $\gamma = -0.5$ and $\omega_2 = 0.65$ we find e.g. $\omega_{1,max} \approx 0.75$. 
\subsection{$\alpha\neq 0$: Interacting boson stars} 
It has been observed in \cite{kk1} that for a single boson star
the space-time becomes flat when the maximal value of $\omega$ is reached 
(consistent with the fact that the boson star vanishes and $Q\rightarrow 0$ in this limit). 
In contrast, space-time become strongly curved close to the origin when the minimal
value of $\omega$ is approached. This feature is correlated to the fact that for small $\omega$ the
matter fields become more concentrated in the region around the origin.

We have studied the domain of existence for $\alpha=0.1$. 
This is given in Fig. \ref{domain_00_11} for
$\gamma=0$ (domain limited by lines with red bullets) and for $\gamma = -1$ (domain limited by lines with blue squares), respectively.
The numerical integration becomes very difficult whenever approaching the limits of  these domains of existence.
The figure demonstrates that both gravity and the direct potential interaction leads to the existence of solutions with lower values
of $\omega_1$.
We find that the interacting solutions stop to exist because the fields $\phi_1$ vanishes identically
when approaching a minimal value of $\omega_2$ (and $\omega_1$ fixed). The other regions of the domain
suggest the emergence of new branches of solutions which backbend from the branches we have constructed.

To understand the features in more detail, we have studied the dependence of the properties
of the solutions on the gravitational coupling $\alpha$ for $\gamma=0$, i.e.
we let the boson stars interact via gravity only. 
We have chosen the frequency of the axially symmetric boson star to be fixed, $\omega_2=0.8$, and have investigated
the properties of the boson stars for varying $\alpha$ and frequency $\omega_1$ of the spherically shaped boson star. As discussed in more detail below, we observe that the number
of solutions available depends strongly on $\omega_1$ and that -- in fact -- additional branches
of solutions exist.

\begin{figure}[!htb]
\centering
\leavevmode\epsfxsize=10.0cm
\epsfbox{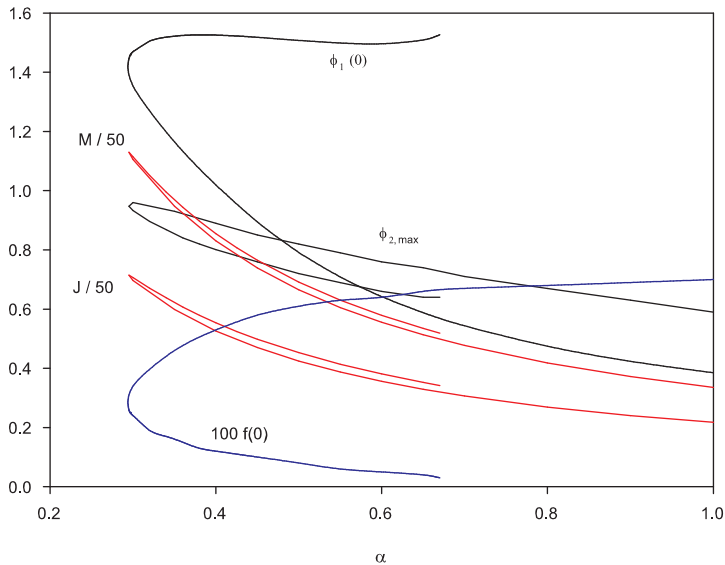}\\
\caption{\label{alpha_vary_02} 
The mass $M$, the angular momentum $J$, the values $f(0)$ and $\phi_1(0)$ as well
as $\phi_{2,max}$ are given in dependence on $\alpha$ for a non-rotating boson star
with $\omega_1=0.2$ interacting with a rotating boson star with $\omega_2=0.8$.}
\end{figure}

We have first studied solutions for which the frequency of the non-rotating
boson star is smaller than that of the rotating boson star. We have chosen $\omega_1=0.2$, $\omega_2=0.8$. For $\alpha=0$, no corresponding
$Q$-ball solution exists. Increasing
$\alpha$ we observe that a pair of solutions exist if $\alpha$ is sufficiently large.
This is illustrated in Fig. \ref{alpha_vary_02}, where we give the mass $M$, the angular momentum $J$,
$f(0)$, $\phi_1(0)$ and $\phi_{2,max}$ in dependence on $\alpha$.
Clearly, two branches of solutions are present, which meet at $\alpha=\alpha_{min}\approx 0.3$, such
that for $\alpha < \alpha_{min}$ no solutions exist.
The branch of solutions with lower mass likely exists for $\alpha \to \infty$, where the scalar fields slowly converge to $\phi_1 = \phi_2 \equiv 0$. The branch with higher energy (we call it ``the second branch'') ceases to
exist at a critical value of $\alpha=\alpha_{cr}$, where $f(0)$ is very small. 
Likely, further branches exist that start at the end point of this second branch. On these
we would expect $f(0)$ to decrease even further towards zero.  
\label{numerics}
\begin{figure}[!htb]
\centering
\leavevmode\epsfxsize=10.0cm
\epsfbox{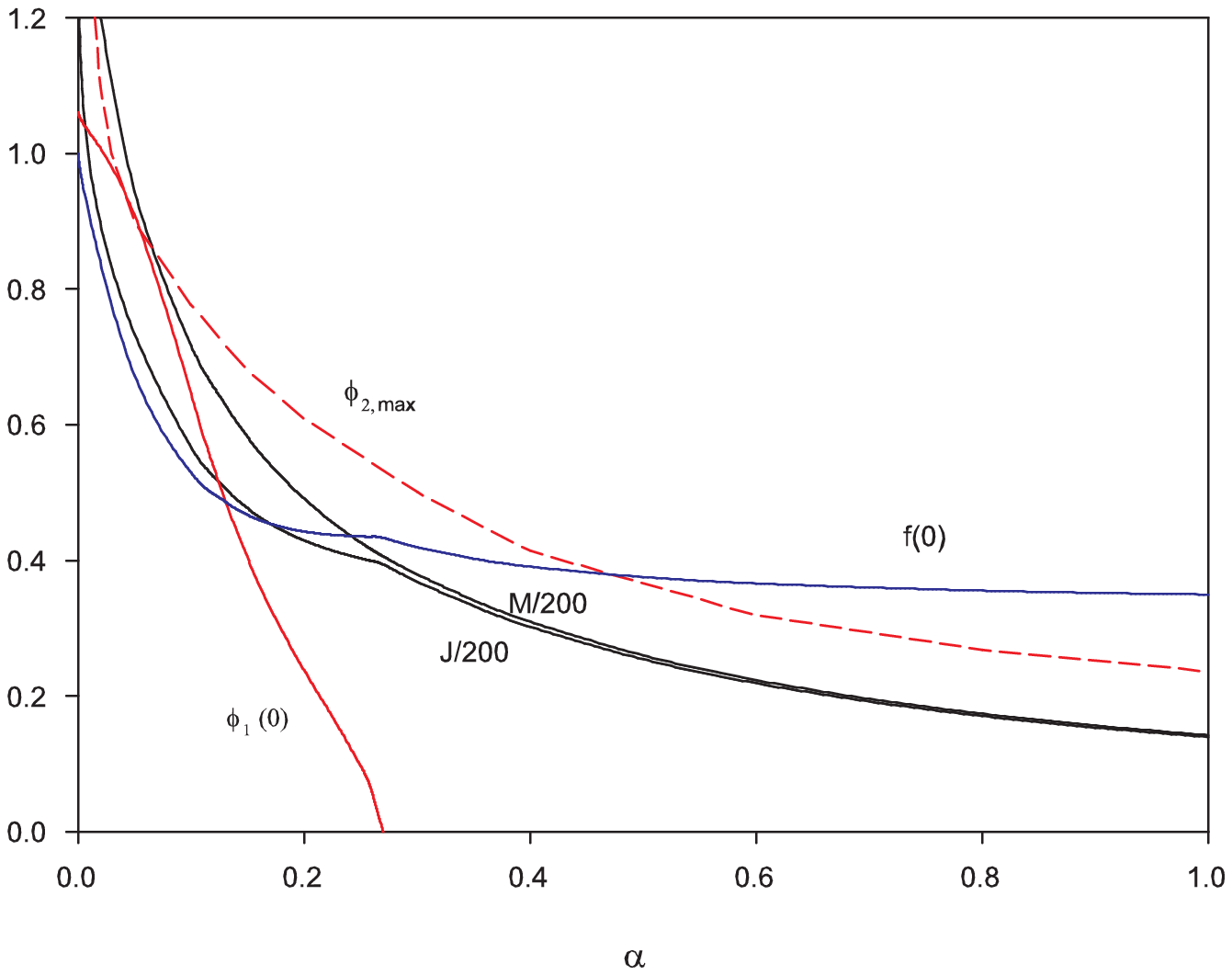}\\
\caption{\label{alpha_vary_08} 
The mass $M$, the angular momentum $J$, the values $f(0)$ and $\phi_1(0)$ as well
as $\phi_{2,max}$ are given in dependence on $\alpha$ for a non-rotating boson star
with $\omega_1=0.8$ interacting with a rotating boson star with $\omega_2=0.8$. Here $\gamma=0$.}
\end{figure}
\label{numerics}
\begin{figure}[!htb]
\centering
\leavevmode\epsfxsize=10.0cm
\epsfbox{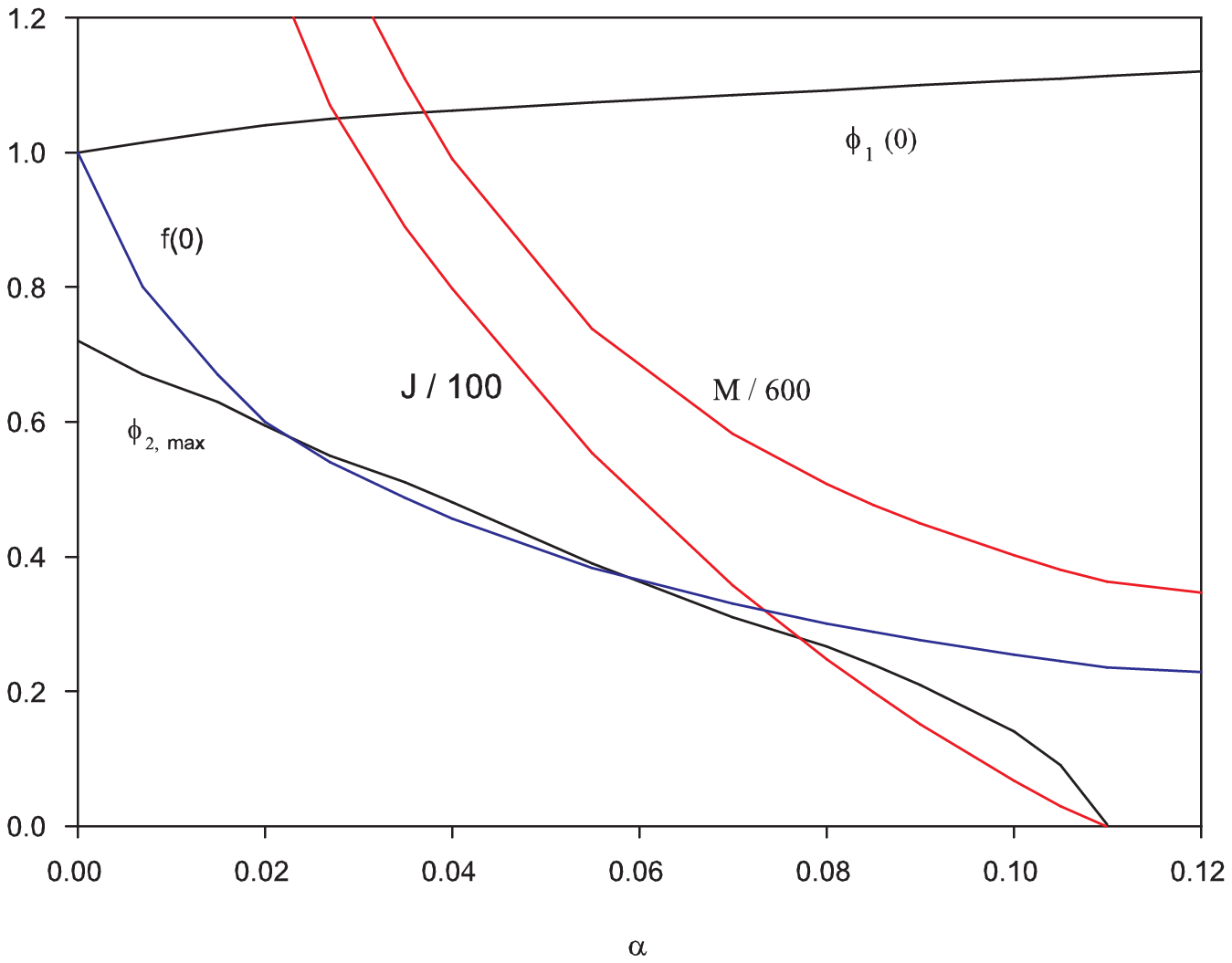}\\
\caption{\label{alpha_vary_04_a} 
The mass $M$, the angular momentum $J$, the values $f(0)$ and $\phi_1(0)$ as well
as $\phi_{2,max}$ are given in dependence on $\alpha$ for a non-rotating boson star
with $\omega_1=0.4$ interacting with a rotating boson star with $\omega_2=0.8$. Here $\gamma=0$.}
\end{figure}
\begin{figure}[!htb]
\centering
\leavevmode\epsfxsize=10.0cm
\epsfbox{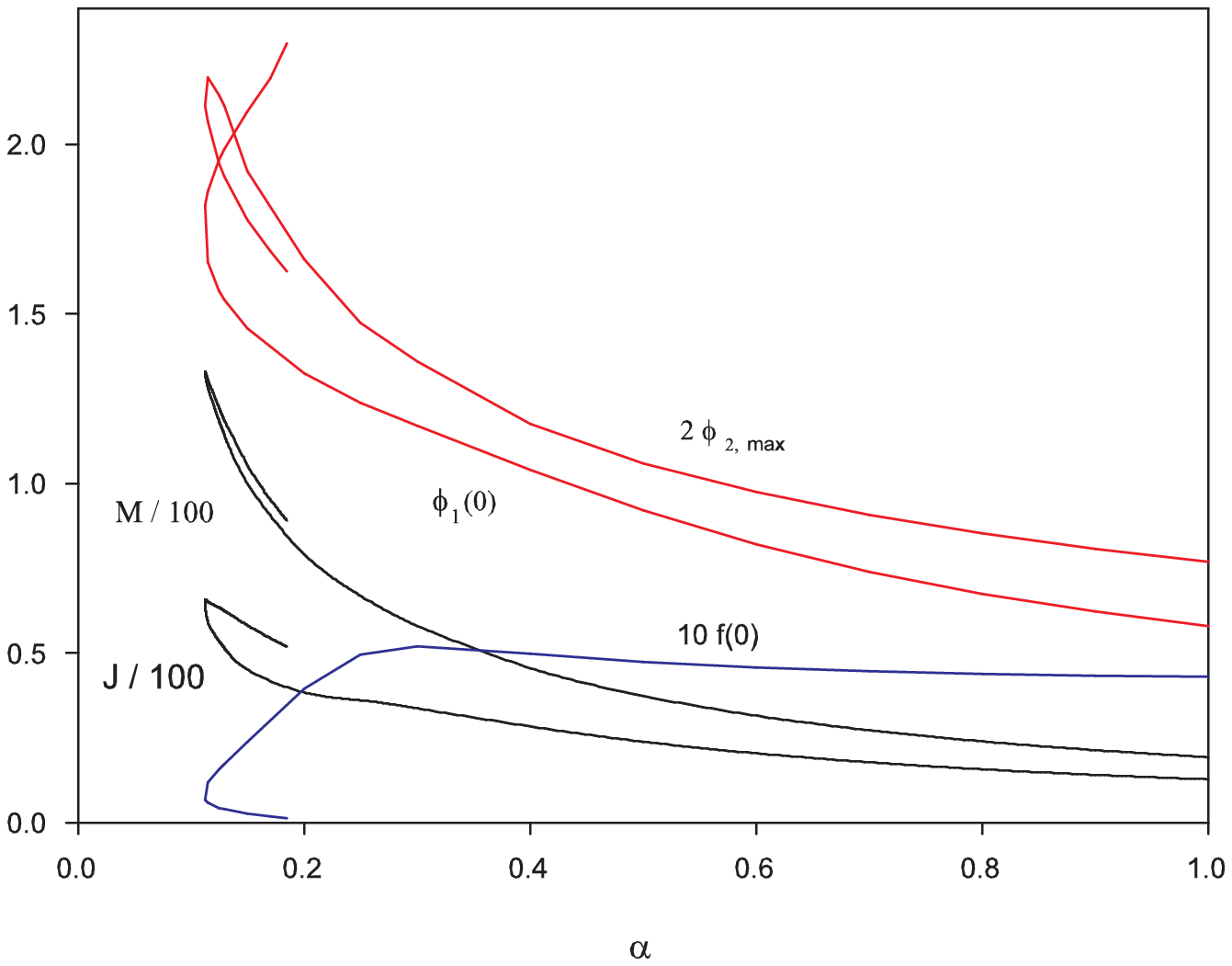}\\
\caption{\label{alpha_vary_04_b} 
The mass $M$, the angular momentum $J$, the values $f(0)$ and $\phi_1(0)$ as well
as $\phi_{2,max}$ are given in dependence on $\alpha$ for a non-rotating boson star
with $\omega_1=0.4$ interacting with a rotating boson star with $\omega_2=0.8$. Here $\gamma=0$. }
\end{figure}

Next, we have studied the case where the frequency of the non-rotating boson star is equal to that of the rotating boson star, i.e. $\omega_2=\omega_2=0.8$. Our results are shown in Fig. \ref{alpha_vary_08}.
In this case, $Q$-ball solutions exist in the flat space-time limit $\alpha=0$. 
For $\alpha > 0$, a single branch of boson stars emerges from these flat space-time solutions.
The mass $M$ and angular momentum $J$ of the interacting boson stars decrease with increasing
gravitational coupling. The values of $\phi_{2,max}$ and $f(0)$ decrease, but stay finite
for all values of the gravitational coupling. However, the value of $\phi_1(0)$ tends to zero
for $\alpha\to \alpha_{cr}\approx 0.24$ indicating that $\phi_1\equiv 0$. This means
that for $\alpha > \alpha_{cr}$ the spherically symmetric boson star has disappeared from the system
and the solution is simply a rotating, single boson star. 

Apparently, the pattern of solutions is quite different for the two cases studied.
To understand the connection between the two patterns, we have studied the solutions
for an intermediate value of $\omega_1$. We have chosen $\omega_2=0.4$ and $\omega_2=0.8$.
We observe that in this case, both patterns of solutions coexist.
For $\alpha=0$ both individual $Q$-balls -- the non-rotating $Q$-ball and the rotating $Q$-ball -- exist and do not interact (remember $\gamma=0$).
Increasing $\alpha$ from zero, these solutions interact via gravity. As is shown in  Fig. \ref{alpha_vary_04_a}, the mass $M$ and angular momentum $J$ decrease for increasing $\alpha$.
For $\alpha\approx 0.110$, however, we find that $J\to 0$ and $\phi_{2,max}\to 0$.
This is related to the fact that in this limit $\phi_2\equiv 0$ such that the rotating
boson star disappears from the system and the solution is simply a single non-rotating boson star.
This is similar to the pattern found for $\omega_1=\omega_2=0.8$ though it is the
rotating boson star that disappears here.

However, we find additional branches for sufficiently large $\alpha$, which are disconnected 
from the $\alpha=0$ solutions.
These branches are shown in Fig. \ref{alpha_vary_04_b}. One branch (the branch with lower energy)
exists up to $\alpha\to \infty$, while the second branch (the one with higher energy)
emerges from the first branch at
$\alpha > 0.113$ and extends back in $\alpha$. On this second branch, the value of $f(0)$ decreases
with increasing $\alpha$ such that at some critical value of $\alpha$, $f(0)$ becomes very small.
Likely further branches exist on which $f(0)$ tends to zero.
This behaviour is similar to that observed for $\omega_1=0.2$ and $\omega_2=0.8$.

To summarize, we find at least three branches of interacting solutions in this case:  
 one branch that is directly connected to the flat space-time limit and exists  for $\alpha \in [0, 0.110]$,
a second branch that exists for $\alpha \in [0.113,\infty]$ and a third branch that exists for
 $\alpha \in [0.113, 0.185]$. The second and third branch join at $\alpha = 0.113$, while the first
branch seems disconnect from the others.

Studying the equations for large values of the gravitational coupling $\alpha$
therefore suggests the following properties:
(i) the interacting solutions obtained by deforming the
flat space-time solutions via gravity do not stay bounded for large $\alpha$ and end up
into single boson stars, where one of the two scalar fields vanishes identically, 
(ii) branches of solutions that have no flat space-time limit do not show the behaviour mentioned
under (i) for large $\alpha$. These branches can somehow be seen as non-perturbative since
they do not exist for arbitrarily small values of $\alpha$.

\section{Conclusions}
We have studied interacting $Q$-balls and boson stars, respectively, in great detail.
While previous work \cite{bh,bh2} has focused on the different types of solutions
possible, we have studied the parameter dependence of the mass and charges of the
solutions here.

We observe new features in the model. In the flat space-time limit the model
describes interacting $Q$-balls. We observe that when $Q$-balls are repelling,
we can make the corresponding frequencies and charges arbitrarily small.
This phenomenon appears for the choices of potential parameters done in previous
work \cite{vw,kk1}. We would like to emphasize that single $Q$-balls with arbitrarily small
charges are also possible, but for different choices of the potential parameters than those
done here and in \cite{vw,kk1}.

Whenever $Q$-balls and boson stars, respectively are interacting, a new type of
solution is possible: one for which $Q_1\neq Q_2$ while $\omega_1=\omega_2$.
We interpret this as a symmetry breaking in the model.
In the flat space-time limit, this is only possible when the $Q$-balls are repelling.
Moreover, asymmetric solutions always have a companion solution in the form of a symmetric
solution with $Q_1=Q_2$. The asymmetric solution has much higher energy 
than the symmetric solution. In a concrete physical setting, we would
thus expect the symmetric solution to be the generic solution.

This changes, however, when boson stars are interacting solely via gravity.
In this case, there is a frequency range in which only asymmetric solutions exist.
This is an important observation when considering boson stars as dark matter candidates.


\section{Appendix: The equations of motion}
Using suitable combinations of the Einstein equations we find the following expressions:
\begin{eqnarray}
 r^2 \partial_{rr} f + \partial_{tt} f &=& -\frac{1}{2}\frac{1}{fl}\left(-2\sin^2\theta l^2 m^2+ 4r f l \partial_r f +
2\frac{\cos\theta}{\sin\theta} f l \partial_{\theta} f + f \partial_{\theta} l \partial_{\theta} f \right.
\nonumber \\
&-& \left.
2 l (r^2 (\partial_r f)^2 + (\partial_{\theta} f)^2) + r^2 f \partial_r l \partial_r f + 4r\sin^2 \theta l^2 m \partial_r m  \right.\nonumber \\
&-& \left. 2\sin^2\theta l^2 ((\partial_r m)^2 + (\partial_{\theta} m)^2)\right) \nonumber \\
&+& \alpha \left(   \frac{glr^2}{f}  T_{tt}
+ f r^2   T_{rr}+ f       T_{\theta\theta}
+ g\left( \frac{m^2 l}{f} + \frac{f}{\sin^2 \theta}\right) T_{\varphi \varphi}
              - 2\frac{g m lr}{f}  T_{t\varphi} \right)
\end{eqnarray}

\begin{eqnarray}
r^2 \partial_{rr} l +   \partial_{tt} l &=& -\frac{1}{2}\frac{1}{fl}\left(6r f l \partial_r l + 4 \frac{\cos\theta}{\sin\theta} f l \partial_{\theta} l - f (r^2 (\partial_r l)^2 + (\partial_{\theta} l)^2)\right) \nonumber \\
& + & \alpha 2 l (r^2 T_{rr} + T_{\theta \theta})
\end{eqnarray}
\begin{eqnarray}
r^2 \partial_{rr} m + \partial_{tt} m &=&  -\frac{1}{2}\frac{1}{fl}\left(-4 fl m + 4rfl \partial_r m
+ 6 \frac{\cos\theta}{\sin\theta} fl \partial_{\theta} m - 4l(r^2 \partial_r f \partial_r m \right. \nonumber \\
&+& \left. \partial_{\theta} f \partial_{\theta} m) + 3f(r^2 \partial_r l \partial_r m + \partial_{\theta} l \partial_{\theta} m) + 4r l m \partial_r f - 3r f m \partial_r l\right)\nonumber \\
& + & \alpha \frac{2g}{\sin^2 \theta}(m T_{\varphi \varphi}-r T_{t \varphi})
\end{eqnarray}

\begin{eqnarray}
     r^2\partial_{rr}g + \partial_{\theta\theta} g&=& \frac{3}{2} \frac{gl}{f^2} 
     \sin^2\theta    \biggl(  r^2 (\partial_r m)^2 + (\partial_{\theta} m)^2 + m^2 - 2 m \partial_r m \biggr) \nonumber \\
     &-& \frac{1}{2f^2} \biggl(  r^2 (\partial_r f)^2 + (\partial_{\theta} f)^2           \biggr) 
     + \frac{1}{2l^2} \biggl(  r^2 (\partial_r l)^2 + (\partial_{\theta} l)^2           \biggr) \nonumber \\    
     &+& \frac{2g}{l}\biggl(  r \partial_r l + \frac{\cos \theta}{\sin \theta} \partial_{\theta} l  \biggr)
     + \frac{1}{g}\biggl( r^2(\partial_r g)^2 + (\partial_{\theta} g)^2 -r g \partial_r g     \biggr) \nonumber \\
&+& \alpha (-2 g r^2 T_{rr} - 2 g T_{\theta\theta}
                           + \frac{2g^2}{\sin^2 \theta} T_{\varphi \varphi})
\end{eqnarray}
where the non-vanishing components of the energy-momentum tensor read:
\begin{eqnarray}
T_{tt} &=& f V(\phi_1,\phi_2) - \sin^2 \theta \frac{l}{f} m^2 V(\phi_1,\phi_2) + \omega_1^2 \phi_1^2+ \omega_2^2 \phi_2^2 - \frac{2}{r} m (k_1 \omega_1 \phi_1^2 + k_2 \omega_2 \phi_2^2) \nonumber\\
&+& \frac{1}{r^2\sin^2\theta} \frac{f^2}{l} (k_1^2 \phi_1^2 + k_2^2 \phi_2^2) + \frac{2}{r} \sin^2\theta \frac{l}{f^2} m^3 (k_1 \omega_1 \phi_1^2 + k_2 \omega_2 \phi_2^2) + \sin^2\theta\frac{l}{f^2} m^2 (\omega_1^2 \phi_1^2 + \omega_2^2 \phi_2^2) \nonumber \\
&-& \frac{2}{r}(\omega_1^2 k_1^2 \phi_1^2 + \omega_2^2 k_2^2 \phi_2^2) + \frac{1}{r^2} \sin^2\theta \frac{l}{f^2}
m^4(k_1^2 \phi_1^2 + k_2^2 \phi_2^2)  
+ \frac{f^2}{lg} ((\partial_r\phi_1)^2 +  (\partial_r\phi_2)^2) \nonumber \\
&-&\sin^2\theta \frac{1}{g} m^2 ((\partial_r\phi_1)^2 +  (\partial_r\phi_2)^2) + \frac{1}{r^2} \frac{f^2}{lg} ((\partial_{\theta}\phi_1)^2 +  (\partial_{\theta}\phi_2)^2) \nonumber\\
&-& \frac{1}{r^2} \sin^2\theta \frac{m^2}{g}( (\partial_{\theta}\phi_1)^2 +  (\partial_{\theta}\phi_2)^2) 
\end{eqnarray}
\begin{eqnarray}
 T_{rr} &=& -\frac{lg}{f} V(\phi_1,\phi_2) + \frac{lg}{f^2} \omega_1^2 \phi_1^2 + \omega_2^2 \phi_2^2 + \frac{2}{r} \frac{lg}{f^2} m (k_1\omega_1 \phi_1^2 + k_2 \omega_2 \phi_2^2) - \frac{g}{r^2 \sin^2\theta}  (k_1^2 \phi_1^2 + k_2^2 \phi_2^2) \nonumber \\
&+& \frac{1}{r^2} \frac{lg}{f^2} m^2 (k_1^2 \phi_1^2 + k_2^2 \phi_2^2) + (\partial_r \phi_1)^2 + (\partial_r \phi_2)^2 - \frac{1}{r^2} (\partial_{\theta} \phi_1)^2 - \frac{1}{r^2} (\partial_{\theta} \phi_2)^2
\end{eqnarray}
\begin{eqnarray}
 T_{\theta\theta}&=& -r^2 \frac{lg}{f} V(\phi_1,\phi_2) + r^2 \frac{lg}{f^2} (\omega_1^2 \phi_1^2 + \omega_2^2 \phi_2^2)
+ 2r \frac{lg}{f^2} m (\omega_1 k_1 \phi_1^2 + \omega_2 k_2 \phi_2^2) \nonumber \\
&+ & \frac{lg}{f^2} m^2 (k_1^2 \phi_1^2 + k_2^2 \phi_2^2) - \frac{g}{\sin^2\theta}(k_1^2 \phi_1^2 + k_2^2 \phi_2^2) \nonumber \\
&-&r^2 (\partial_r\phi_1)^2 -r^2 (\partial_r\phi_2)^2 + (\partial_{\theta}\phi_1)^2+ (\partial_{\theta}\phi_2)^2
\end{eqnarray}
\begin{eqnarray}
 T_{r\theta} = 2 (\partial_r \phi_1)(\partial_{\theta} \phi_1) + 2 (\partial_r \phi_2)(\partial_{\theta} \phi_2) 
\end{eqnarray}
\begin{eqnarray}
 T_{t\varphi} &=& r \sin^2\theta \frac{l}{f} m V(\phi_1,\phi_2) - r\sin^2\theta \frac{l}{f^2} m(\omega_1^2 \phi_1^2 + \omega_2^2 \phi_2^2) - 2\sin^2\theta \frac{l}{f^2} m^2 (k_1\omega_1 \phi_1^2 + k_2\omega_2 \phi_2^2) \nonumber \\
& + & \frac{1}{r} m (k_1^2 \phi_1^2 + k_2^2 \phi_2^2) - \frac{1}{r}\sin^2\theta \frac{l}{f^2} m^3 (k_1^2 \phi_1^2 + k_2^2 \phi_2^2) +2 k_1 \omega_1 \phi_1^2 + 2 k_2 \omega_2 \phi_2^2 \nonumber \\
&+& r\sin^2\theta \frac{m}{g} ((\partial_r \phi_1)^2 + (\partial_r\phi_2)^2) + \frac{\sin^2\theta}{r} \frac{1}{g} ((\partial_{\theta} \phi_1)^2 + (\partial_{\theta}\phi_2)^2)
\end{eqnarray}
\begin{eqnarray}
 T_{\varphi\varphi} &=& -r^2 \sin^2\theta \frac{l}{f}V(\phi_1,\phi_2) + r^2 \sin^2\theta \frac{l}{f^2} (
\omega_1^2 \phi_1^2 + \omega_2^2 \phi_2^2) + 2r\sin^2\theta \frac{l}{f^2} m (k_1\omega_1 \phi_1^2 + 
k_2\omega_2 \phi_2^2) \nonumber \\
&+& k_1^2 \phi_1^2 + k_2^2 \phi_2^2  + \sin^2\theta \frac{l}{f^2} m^2 (k_1^2\phi_1^2 +k_2^2\phi_2^2) - r^2 \sin^2\theta \frac{1}{g}((\partial_r\phi_1)^2 +(\partial_r \phi_2)^2) \nonumber \\
&-& \sin^2\theta \frac{1}{g} ((\partial_{\theta}\phi_1)^2 + (\partial_{\theta} \phi_2)^2)
\end{eqnarray}
Finally, the Euler-Lagrange equations read ($i=1,2$):
\begin{eqnarray}
 r^2 \partial_{rr} \phi_i + \partial_{\theta\theta} \phi_i &=&
-\frac{1}{2}\frac{1}{fl}\left(-r^2 l^2 g \frac{\partial V}
{\partial \phi_i} - 2 \frac{k_i^2}{\sin^2\theta} 
f l g \phi_i
+  2 \frac{l^2 g }{f} (m k_i + r \omega_i)^2
\phi_i \right. \nonumber \\
&+& \left. 4rfl \partial_r \phi_i 
+ 2\frac{\cos\theta}{\sin\theta} f l \partial_{\theta} \phi_i +
f (r^2 \partial_r l \partial_r \phi_i + \partial_{\theta} l \partial_{\theta}\phi_i)\right)
\end{eqnarray}

\end{document}